\documentclass[12pt,preprint]{aastex}

\received{}
\accepted{}

\newcommand{\AAA}{$\rm{\AA} \;$}
\newcommand{\kms}{km~s$^{-1}$}
\newcommand{\kmss}{km~s$^{-1} \;$}

\author{
Timothy A. Reichard,\altaffilmark{1}
Gordon T. Richards,\altaffilmark{1,2}
Donald P. Schneider,\altaffilmark{1}
Patrick B. Hall,\altaffilmark{2,3}
Alin Tolea,\altaffilmark{4}
Julian H. Krolik,\altaffilmark{4}
Zlatan Tsvetanov,\altaffilmark{4}
Daniel E. Vanden Berk,\altaffilmark{5}
Donald G. York,\altaffilmark{6,7}
G.R. Knapp,\altaffilmark{2}
James E. Gunn,\altaffilmark{2}
and J. Brinkmann\altaffilmark{8}
}

\altaffiltext{1}{Department of Astronomy and Astrophysics, The Pennsylvania State University, 525 Davey Laboratory, University Park, PA 16802.}
\altaffiltext{2}{Princeton University Observatory, Peyton Hall, Princeton, NJ 08544.}
\altaffiltext{3}{Departamento de Astronom\'{\i}a y Astrof\'{\i}sica, Pontificia Universidad Cat\'{o}lica de Chile, Casilla 306, Santiago 22, Chile.}
\altaffiltext{4}{Department of Physics and Astronomy, The Johns Hopkins University, 3400 North Charles Street, Baltimore, MD 21218-2686.}
\altaffiltext{5}{Department of Physics and Astronomy, University of Pittsburgh, 3941 O'Hara Street, Pittsburgh, PA 15260.}
\altaffiltext{6}{Department of Astronomy and Astrophysics, The University of Chicago, 5640 South Ellis Avenue, Chicago, IL 60637.}
\altaffiltext{7}{Enrico Fermi Institute, The University of Chicago, 5640 South Ellis Avenue, Chicago, IL 60637.}
\altaffiltext{8}{Apache Point Observatory, P.O. Box 59, Sunspot, NM 88349.}

\begin{document}

\title{A Catalog of Broad Absorption Line Quasars from the Sloan Digital
Sky Survey Early Data Release}

\begin{abstract}

We present a catalog of 224 broad absorption line quasars (BALQSOs)
from the Sloan Digital Sky Survey's Early Data Release Quasar Catalog,
including a relatively complete and homogeneous subsample of 131
BALQSOs.  Since the identification of BALQSOs is subject to
considerable systematic uncertainties, we attempt to create a complete
sample of SDSS BALQSOs by combining the results of two automated
selection algorithms and a by-eye classification scheme.  One of these
automated algorithms finds broad absorption line troughs by comparison
with a composite quasar spectrum.  We present the details of this
algorithm and compare this method to that which uses a power-law fit
to the continuum.  The BALQSOs in our sample are further classified as
high-ionization BALQSOs (HiBALs), low-ionization BALQSOs (LoBALs), and
BALQSOs with excited iron absorption features (FeLoBALs); composite
spectra of each type are presented.  We further present a study of the
properties of the BALQSOs in terms of the balnicity distribution,
which rises with decreasing balnicity.  This distribution of
balnicities suggests that the fraction of quasars with intrinsic
outflows may be significantly underestimated.

\end{abstract}

\keywords{quasars: general --- quasars: absorption lines}

\section{Introduction}

The nature of BALQSOs has been a question at the forefront of quasar
research for the past two decades \markcite{tur84,wey95}({Turnshek} 1984; {Weymann} 1995), and
considerable uncertainty remains about the nature of the absorption.
One proposed explanation of the broad absorption line (BAL) phenomenon
is that BALQSOs and nonBALQSOs are distinct populations of objects
\markcite{sh87}({Surdej} \& {Hutsemekers} 1987).  Similarly some have argued that only LoBALs are a
different class of quasars \markcite{bm92}({Boroson} \& {Meyers} 1992), while others suggest that
BALQSOs and nonBALQSOs are the same type of quasar but viewed from
different orientations \markcite{wmf+91,ocm+99,sh99}({Weymann} {et~al.} 1991; {Ogle} {et~al.} 1999; {Schmidt} \& {Hines} 1999) or at different
stages in their life cycles \markcite{bwg+00}({Becker} {et~al.} 2000).

Regardless of the nature of their hosts, it is clear that BAL troughs
are caused by outflowing gas that is intrinsic to the quasar and are
not produced by galaxies along the line of sight (as is the case for
most narrow absorption systems).  The traditional definition of a BAL
trough requires that the outflows extend to considerable velocities
from the quasar's emission redshift; however, it is possible that
these intrinsic outflows extend to smaller velocities and are related
to the so-called ``associated'' absorbers that are not as broad and
have smaller terminal velocities \markcite{fwp+86}({Foltz} {et~al.} 1986).  Some of these
associated systems are known to be intrinsic outflows, but others may
simply be the result of absorption in the host galaxy or a nearby
galaxy.

The characteristics of BALQSOs have been the subject of a number of
previous studies.  Working from the Large Bright Quasar Survey (LBQS:
\markcite{hfc+95}{Hewett}, {Foltz}, \& {Chaffee} 1995), \markcite{wmf+91}{Weymann} {et~al.} (1991) compiled the largest previous BALQSO
sample.  They found that, after approximately correcting for selection
effects, BALs occurred in $\simeq 12\%$ of all quasars, but with a
striking lack of BALs in radio-loud quasars.  They also created
composite spectra for both BAL and ``normal" quasars, using 42
nonBALQSOs and 40 BALQSOs, including 6 low-ionization BALQSOs (LoBALs)
and 34 high-ionization BALQSOs (HiBALs).  More recently \markcite{btb+01}{Brotherton} {et~al.} (2001)
created composite BALQSO spectra using 25 HiBALs and 18 LoBALs from
the FIRST Bright Quasar Survey (FBQS; \markcite{wbg+00}{White} {et~al.} 2000).

The Sloan Digital Sky Survey (SDSS; \markcite{yor+00}{York} {et~al.} 2000) will produce a
sample of quasars that is more than 100 times larger than either the
LBQS or FBQS, and the SDSS should be relatively unbiased with respect
to BALQSOs.  Already, \markcite{mvi+01}{Menou} {et~al.} (2001) have examined 13 radio-detected
SDSS/FIRST quasars with BAL-like intrinsic absorption.  The properties
of the most extreme cases of SDSS BALQSOs have been presented by
\markcite{hal+02}{Hall} {et~al.} (2002).  \markcite{kro+02}{Tolea}, {Krolik}, \& {Tsvetanov} (2002) discussed the principal statistical
properties of a sample of \ion{C}{4} BALs drawn from a restricted
subset of the EDR quasars studied herein.

The size of the sample of BALQSOs produced by the SDSS suggests that
the time is ripe to explore new methods for identifying outflows in
AGN.  \markcite{hal+02}{Hall} {et~al.} (2002) have suggested a modification for the
classification of BALQSOs that imposes less restrictive constraints on
the extent of the absorption in BALQSOs.  Herein, we explore a
different method of determining the underlying continuum and emission
line flux, and compare the results with the standard BAL approach as
was used by \markcite{kro+02}{Tolea} {et~al.} (2002).

Thus, the primary goals of this new paper are three-fold: (1) the
comparison of our composite based continuum+emission line fitting
procedure with the traditional power-law continuum plus Gaussian
emission line fitting procedure for defining BALQSOs, (2) the
construction of a large, well-defined sample of BALQSOs, and (3) the
presentation of the resulting distribution of balnicity indices for
BALQSOs.

The paper by \markcite{kro+02}{Tolea} {et~al.} (2002) and this paper differ in a number of ways
including (1) the definition of the underlying continuum+emission, (2)
the way the sample is restricted in order to produce a more
homogeneous sample, and (3) the selection and classification of
LoBALs.  In this paper we will compare and contrast our results to
\markcite{kro+02}{Tolea} {et~al.} (2002) and show that the two methods/samples produce similar
results.  A detailed investigation of the continuum and emission line
properties of these BALQSOs will be presented by \markcite{rrh+02}{Reichard et al.} (2003).  We
reserve detailed discussion of the BAL fraction as a function of
redshift to \markcite{rrh+02}{Reichard et al.} (2003) since there are color dependent selection
effects that must be considered.

In \S~\ref{sec:data} we define our quasar sample.  A discussion of the
process by which we fit individual spectra to a composite spectrum to
define the intrinsic flux level is given in \S~\ref{sec:cont}.  Our
method of classifying quasars into nonBAL, HiBAL, LoBAL, and FeLoBAL
subsamples can be found in \S~\ref{sec:catalog}, which also presents
our BALQSO catalog and composite spectra of our subsamples.  The
catalog has two parts: a relatively complete subsample of BALQSOs that
meet the traditional BALQSO definition (nonzero balnicity index) and
lie within the redshift range $1.7 \le z \le 4.2$, and a supplementary
listing of BALQSOs largely outside this redshift range.  Using our
BALQSO catalog, we discuss the balnicity index distribution in
\S\ref{sec:discussion}; \S\ref{sec:conclusions} summarizes our
results.  Throughout this paper we will use the cosmology that is
traditionally used to define quasar luminosities, where $H_o
={\rm50\,km\,s^{-1}\,Mpc^{-1}}$, $\Omega_M = 1$, and $\Omega_\Lambda =
0$.  We also adopt a convention for optical spectral index such that
$\alpha=\alpha_{\lambda}$ unless stated otherwise, where $f_{\lambda}
\propto \lambda^{\alpha_{\lambda}}$.

\section{The Data \label{sec:data}}

The parent sample for our investigation of BALQSOs consists of the
3814 bona fide quasars ($M_{i^*}\,<\,-23$, with at least one line
broader than $1000\,{\rm km\,s^{-1}}$) from the SDSS Early Data
Release (EDR; \markcite{sto+02}{Stoughton} {et~al.} 2002) quasar catalog \markcite{sch+02}({Schneider} {et~al.} 2002).  These
quasars were selected for spectroscopic follow-up from the SDSS
imaging survey, which uses a wide-field multi-CCD camera
\markcite{gcr+98}({Gunn} {et~al.} 1998).  The spectra cover the optical range 3800$-$9200 \AAA
at a resolution of 1800$-$2100.  As discussed by \markcite{sto+02}{Stoughton} {et~al.} (2002),
quasar candidates were identified using three different preliminary
versions of the SDSS Quasar Target Selection Algorithm
\markcite{rfn+02}({Richards} {et~al.} 2002a), which identifies quasar candidates according to their
broad-band SDSS colors \markcite{fig+96,sto+02}({Fukugita} {et~al.} 1996; {Stoughton} {et~al.} 2002).  EDR quasars were also
identified as optical matches to both radio sources from the VLA
``FIRST'' survey \markcite{bwh95}({Becker}, {White}, \& {Helfand} 1995) and {\em ROSAT} All-Sky Survey
\markcite{vab+99}({Voges} {et~al.} 1999) sources and also as ``serendipitous'' sources (see
\S~4.8.4.3 in \markcite{sto+02}{Stoughton} {et~al.} 2002).  Details of the SDSS photometric
calibrations are given by \markcite{hfs+01}{Hogg} {et~al.} (2001) and \markcite{stk+02}{Smith} {et~al.} (2002), and the
astrometric calibration is described by \markcite{pmh+02}{Pier et al.} (2002).  The
spectroscopic tiling algorithm is discussed by \markcite{blm+02}{Blanton et al.} (2002).

Although the quasar selection algorithm has been evolving, the
differences between the EDR versions of quasar target selection
\markcite{sto+02}({Stoughton} {et~al.} 2002) and the final version of quasar target selection
\markcite{rfn+02}{Richards} {et~al.} (2002a) are largely for quasars with $z\sim3.5$ and $z\sim4.5$.
Thus the BALQSOs studied herein should be representative of the
BALQSOs that will be discovered as part of the final, homogeneous SDSS
quasar survey (at least for those quasars with $i^*<19.1$ --- the SDSS
quasar target selection magnitude limit; in the present study we also
include some fainter quasars from the EDR quasar sample that were
selected to fainter limits).  In order to define a more homogeneously
selected sample, we have also indicated which of the BALQSOs presented
in this catalog would have been selected using the final SDSS quasar
selection algorithm presented by \markcite{rfn+02}{Richards} {et~al.} (2002a).

\section{Composite Spectra Fitting \label{sec:cont}}

\subsection{Overview}

Determining whether a quasar is a BALQSO is a complicated task.  The
standard method is to calculate the ``balnicity'' index (BI), defined
by \markcite{wmf+91}{Weymann} {et~al.} (1991).\footnote{The BI is essentially a modified
equivalent width in velocity space and is defined as follows.
Absorption between 3000 and 25000 km s$^{-1}$ blueward of C~IV
emission redshift is integrated so long as the absorption falls at
least 10\% below the continuum for at least 2000 km s$^{-1}$.  The
25000 km s$^{-1}$ limit is chosen to avoid emission and absorption
from Si~IV.  Any absorption within $3000\,{\rm km\,s^{-1}}$ and that
fails to span at least $2000\,{\rm km\,s^{-1}}$ in width is excluded
in order to avoid contamination from absorption that might not be due
to an outflow, specifically the so-called ``associated'' absorption
lines \markcite{fwp+86}({Foltz} {et~al.} 1986).  BIs can range from 0 to 20000~km~s$^{-1}$.}  A
BI of zero indicates that broad absorption is absent, while a positive
BI indicates not only the presence of one or more broad absorption
troughs, but also the amount of absorption.  However, the
\markcite{wmf+91}{Weymann} {et~al.} (1991) definition is not free of complications; specifically,
how does one define the true continuum+emission line level and the
systemic redshift when there is significant absorption?  Broad
absorption lines remove flux from the blue wing of emission lines,
thus the flux level from which to measure absorption is not the
continuum alone, but must also include emission line flux.

To measure BIs in our BALQSO sample, we used two largely automated
methods.  One of the automated schemes is described in \markcite{kro+02}{Tolea} {et~al.} (2002).
The other is new, and so we describe it in detail here.

In previous studies, the general procedure has been to approximate the
continuum by a selected analytic form (usually a power-law) and fit it
to the spectrum, avoiding (to the degree possible) those portions with
strong line features, whether emission or absorption.  The emission
line corresponding to BAL features is modeled by mirror-imaging the
profile of the red wing onto the blue side about either the measured
or expected peak of \ion{C}{4}.  This is essentially the process that
was used by \markcite{kro+02}{Tolea} {et~al.} (2002) to measure balnicity indices in their
BALQSO sample of 116 objects.  Although this process works relatively
well, the overall similarity of quasar spectra combined with issues
such as emission line blueshifts and asymmetries suggest that a
template fitting procedure might work as well as the traditional
method.  Thus we have developed an automated procedure based on
fitting to a template quasar spectrum; we will compare the results of
this process to the more traditional method.

From \markcite{rfs+01}{Richards} {et~al.} (2001) and \markcite{vrb+01}{Vanden Berk} {et~al.} (2001), we know that quasars at a
given redshift are very similar in the UV/optical part of the
spectrum.  Although there are small differences in the continuum slope
and the strength of the emission lines, and there is a small fraction
(few percent) of anomalously red quasars in the SDSS magnitude range
\markcite{rhv+02}({Richards et al.} 2003), the average quasar spectrum is quite representative of
the sample as a whole.  On the basis of this perceived similarity of
different quasar spectra, we chose to use the EDR composite quasar
spectrum to define the continuum and emission line levels for our
balnicity index determinations in this new composite-based method.
The EDR composite quasar spectrum was created in the same manner as
the \markcite{vrb+01}{Vanden Berk} {et~al.} (2001) quasar composite spectrum, but using only the 3814
EDR quasars; note that BALQSOs have not been removed from the EDR
composite quasar spectrum --- by definition.

We model the true continuum and emission line levels for a quasar with
unknown properties by fitting the EDR composite quasar spectrum to the
input spectrum that we wish to check for BAL troughs.  In matching the
composite quasar spectrum to the input spectrum, we must allow for
changes in the slope and shape of the continuum between the input
quasar spectrum and the template quasar spectrum.  We adjust the
overall slope by multiplying by a power-law; we model additional
spectral ``curvature" relative to the composite in terms of dust
extinction.  We now turn to a discussion of these issues.

\subsection{Dust Reddening Laws \label{sec:redlaws}}

To allow for changes in the shape of the continuum, we make use of a
dust reddening law with color excess (reddening), $E(B-V)$, as a
parameter in our fits.  Dust reddening is a function of wavelength,
and reddening by dust will introduce curvature into an otherwise
power-law continuum.  We emphasize that there is no a priori reason to
require that the curvature in a spectrum be caused by dust; however,
there is growing evidence that BALQSOs can be dust reddened
\markcite{sf92,yv99,btb+01,hal+02}(e.g., {Sprayberry} \& {Foltz} 1992; {Yamamoto} \& {Vansevi{\v c}ius} 1999; {Brotherton} {et~al.} 2001; {Hall} {et~al.} 2002).  Still, it could be that any
curvature or apparent reddening in the spectrum of a quasar is caused
by processes other than dust; we will elaborate on this issue more in
\markcite{rrh+02}{Reichard et al.} (2003).

For our present purposes, however, we will assume that the cause of
any curvature in the spectrum of a quasar relative to the composite is
due to dust reddening (and extinction).  We will determine the amount
of ``reddening" and power-law spectral index adjustment by
simultaneously fitting for both parameters.  To simplify the process,
we will assume that any dust reddening is located at the quasar
redshift and is not caused by dust exterior to the quasar along our
line of sight.  If the reddening instead occurs primarily along the
line of sight, it will simply change the resulting values of $E(B-V)$,
but will not significantly change the amount of curvature in the
spectrum.

A review of the extinction due to dust is given by \markcite{sm79}{Savage} \& {Mathis} (1979); see
also \markcite{mb81}{Mihalas} \& {Binney} (1981).  The extinction, typically written as $A_{\lambda}$
at a given wavelength, $\lambda$, is given by
\begin{equation}
A_{\lambda} = E(\lambda - V) + R_V\times E(B - V),
\end{equation}
where $E(\lambda - V) \equiv A_\lambda - A_V$ is the color excess.
The observed extinction laws typically have an approximately
$\lambda^{-1}$ dependence, but there are differences among the
extinction laws for the Milky Way \markcite{ccm89}({Cardelli}, {Clayton}, \& {Mathis} 1989), the Large Magellanic
Cloud (LMC; \markcite{nmw+81}{Nandy} {et~al.} 1981), and the Small Magellanic Cloud (SMC;
\markcite{plp+84}{Prevot} {et~al.} 1984).  The main differences are the relative slopes of
the extinction as functions of $\lambda$ and the strength of the
so-called ``$2200\,{\rm \AA}$ bump''; see \markcite{sf92}{Sprayberry} \& {Foltz} (1992).

For most extragalactic sources, including quasars, the $2200\,{\rm
\AA}$ bump is either weak or missing, and the SMC reddening law is
often assumed to be the most appropriate law to use (it has the
weakest $2200\,{\rm \AA}$ bump).  Throughout this paper we will use
the SMC extinction curve as given by \markcite{pei92}{Pei} (1992), which has
$R_V=2.93$ and goes roughly as $\lambda^{-1}$ in the UV/optical part
of the spectrum that is covered by our spectra.

\subsection{Continuum Adjustment\label{sec:fitcomposite}}

During the process of adjusting the continuum of the composite
spectrum to match each input spectrum, we assume that the quasar
continua follow a power-law relation that is diminished by dust
extinction characterized by the color excess $E \equiv E(B - V)$ using
the \markcite{pei92}{Pei} (1992) SMC reddening law.  We begin by expressing a
spectrum as
\begin{equation}
f (\lambda; \alpha, E) \propto \lambda^{\alpha} 10^{-a E \xi (\lambda)},
\end{equation}
where $a = 0.4 (1+R_V)$, $\xi (\lambda)$ is the extinction curve as
given by \markcite{pei92}{Pei} (1992), and rest wavelengths, $\lambda$, are given in
microns.  We determine the spectral index, $\alpha$
$(=\alpha_\lambda=-2-\alpha_\nu)$, and reddening, $E$, of each input
spectrum by fitting to each an adjusted composite spectrum.  The
composite spectrum $f_c (\lambda; \alpha_c, E_c)$ is adjusted to
$f_c^\prime (\lambda; \alpha, E)$ by changing its spectral index and
reddening using
\begin{equation} f_c^\prime (\lambda; \alpha, E) = f_c (\lambda;
\alpha_c, E_c) \lambda^{(\alpha - \alpha_c)} 10^{-a(E -
E_c) \xi(\lambda)}.
\end{equation}
We fit the composite spectrum to each input spectrum by minimizing a
weighted $\chi^2$ function.  Minimization is more easily achieved by
smoothing the input spectra before evaluating the $\chi^2$ function,
and so we smooth by 15 pixels.  Expressing an adjusted composite
spectrum as $f_c^\prime (\lambda; \alpha, E)$ and a smoothed input
spectrum as $f_i (\lambda; \alpha_i, E_i)$ with error spectrum
$\sigma_i (\lambda)$, we find the values of $\alpha$ and $E$ that
minimize the $\chi^2$ function
\begin{equation}
\chi^2 (\alpha, E) = \frac{\sum_\lambda \left (\frac{f_c^\prime (\lambda; \alpha, E) - f_i (\lambda; \alpha_i, E_i)}{\sigma _i (\lambda)} \right )^2 w(\lambda)}{\sum_\lambda w(\lambda)}.
\end{equation}

\begin{figure}[p]
\epsscale{1.0}
\plotone{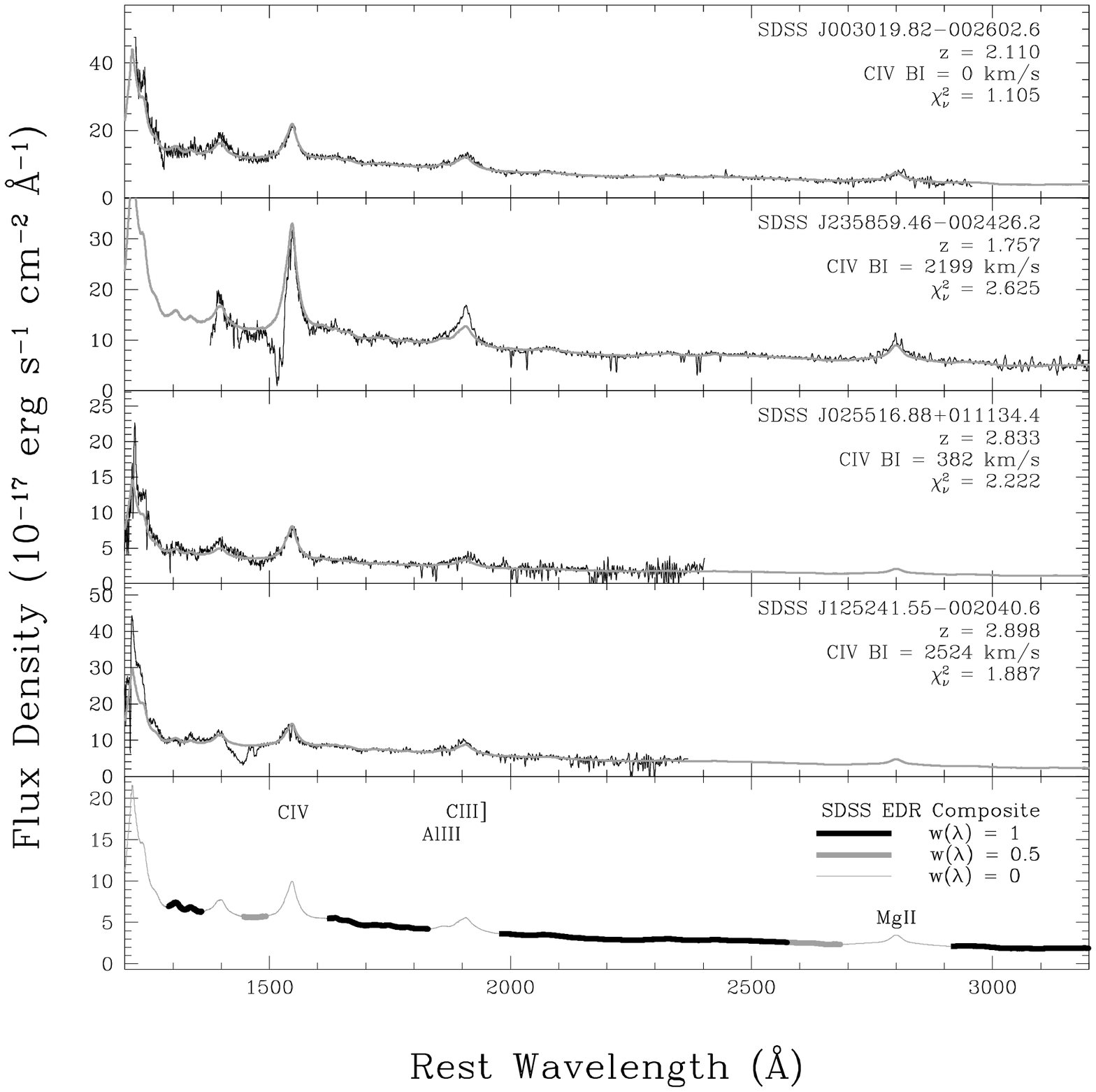}
\caption{EDR composite spectrum fits ({\em gray}) for four quasar
spectra ({\em black}) from the EDR sample: one nonBAL ({\em top
panel}) and three HiBALs ({\em middle three panels}).  Note how well
the composite fitting method can recover BALQSOs with very weak
absorption troughs such as the BALQSO shown in the third panel.  The
EDR composite spectrum is fitted by a power law with SMC reddening.
The C IV emission line of the EDR composite has been scaled to match
the peak line flux of each object.  The spectra are normalized at
$1725(1 + z)$ \AA.  The last panel shows the weights as a function of
wavelength that were used during our composite fitting procedure; see
text for more details. \label{fig:fig1}}
\end{figure}

The weight function $w(\lambda)$ is used to remove or reduce the
effects of prominent emission lines.  The emission line regions
excised are the lower and upper wavelength limits given by
\markcite{vrb+01}{Vanden Berk} {et~al.} (2001) for Ly$\alpha$+\ion{N}{5}, \ion{Si}{4}, \ion{C}{4},
\ion{C}{3}], and \ion{Mg}{2}.  The other excluded and deweighted
regions are the \ion{C}{4} broad absorption line region (inclusive
from the \ion{Si}{4} emission line peak to the \ion{C}{4} peak), the
\ion{Mg}{2} broad absorption line region (from $\lambda2575\,{\rm
\AA}$ to $\lambda2686\,{\rm \AA}$), the Lyman-$\alpha$ forest
(shortward of $\lambda1050\,{\rm \AA}$), and wavelengths redward of
4150 \AAA (the longest wavelength at which the composite spectrum
matches a power law continuum line, as shown in Figure 6 of
\markcite{vrb+01}{Vanden Berk} {et~al.} 2001).  For a pixel in one of the broad absorption
regions, $w(\lambda) = 0.5$; in one of these other excluded wavelength
ranges, $w(\lambda) = 0$; otherwise, $w(\lambda) = 1$.  These regions
are depicted in the bottom panel of Figure~\ref{fig:fig1}.  The weight
function ensures that the continua of the composite and input spectra
are matched without having emission and absorption regions affecting
the spectral index and reddening determinations.  Although we would
like to completely discount the broad absorption regions during the
fitting process, fits are improved by giving the regions a lower,
nonzero weight where the broad absorption region is the bluest
accessible part of the spectrum.  For example, in fitting the
composite spectrum to compute a \ion{C}{4} balnicity index for an
object with $z \sim 1.70$, the \ion{C}{4} broad absorption region must
be given some weight; otherwise, little of the spectrum blueward of
the red tail of the \ion{C}{4} emission line is counted, and the
fitted composite may diverge blueward of the emission line.

Before the $\chi^2$ function is evaluated, the adjusted composite
spectrum is normalized to match the average flux density of the input
spectrum in a small wavelength range.  We chose different
normalization windows when defining a continuum for \ion{C}{4} and
\ion{Mg}{2} balnicity indices: $1725 \pm 25$ \AAA for \ion{C}{4} ($1.7
\le z \le 4.2$), and $3150 \pm 25$ \AAA $(0.5 \le z \le 1.9)$ and
$2200 \pm 25$~\AAA $(1.9 < z \le 2.1)$ for \ion{Mg}{2}.  The 3150 \AAA
window is preferred for computing \ion{Mg}{2} balnicity indices
because of its close proximity to the \ion{Mg}{2} emission line, but
it is redshifted beyond the ends of the good spectral range for SDSS
spectra with redshifts greater than $\sim$1.9.  We choose 2200 \AAA
for higher redshifts ($1.9 < z \le 2.1$) for which the \ion{Mg}{2}
absorption region remains in the spectral range.  These values are
chosen to be in wavelength ranges that are accessible for all of the
objects in the redshifted wavelength ranges considered and that are
reasonable local continua (i.e., not affected by strong emission or
absorption lines); see \S\ref{sec:catalog}.

The $\chi^2$ function is minimized by employing a modified
Newton-Raphson method \markcite{ptv+92}({Press} {et~al.} 1992), which quadratically converges to
the values of $\alpha$ and $E$ that minimize the function.  Using
$\alpha = \alpha_c$ and $E = E_c$ as initial guesses yields a
convergence fraction of 92\%.  When convergence succeeds, the $\chi^2$
function is evaluated at points $(\alpha, E)$ close to the convergence
point $(\alpha_i, E_i)$, and the convergence point is adjusted as
necessary if a new local minimum is found.  When convergence fails, we
evaluate the $\chi^2$ function over a wide grid of $(\alpha, E)$
points and choose the minimizing values.  Despite the simplicity of
this latter approach, we prefer the Newton-Raphson method as the
primary method because of its fast convergence.

Examples of the results of this procedure are shown in
Figure~\ref{fig:fig1}.  Here we show the continuum-adjusted
composite quasar spectra (gray line) over-plotted on a sample
nonBALQSO (top panel) and three HiBALs (middle three panels).  The
\ion{C}{4} emission lines of the EDR composite spectrum have been
scaled to match the peak line flux of the objects; see
\S\ref{sec:empeak}.  The spectra are normalized at the redshifted
wavelength $1725(1+z)\,{\rm \AA}$.

\subsection{Goodness of Fit Analysis}

\begin{figure}[p]
\epsscale{1.0}
\plotone{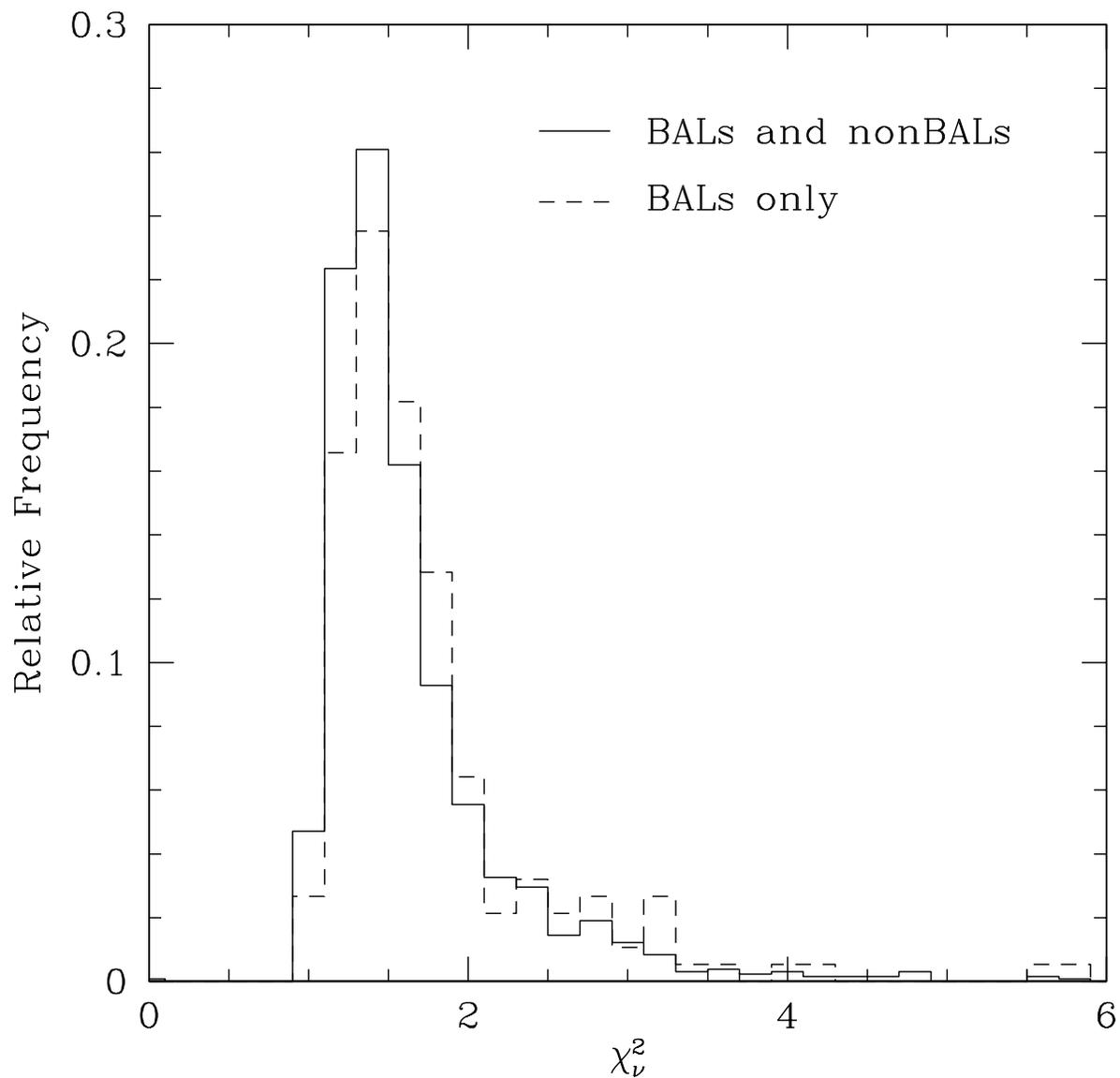}
\caption{Distribution of $\chi^2_{\nu}$ values for all the objects in
the redshift range searched for BAL troughs ({\em solid line}) and for
all quasars that we have classified as BALQSOs ({\em dashed line}).
Quasars with $\chi^2_{\nu} \lesssim 2.5$ are considered to have good
fits.  Larger values indicate either bad fits, or significant
absorption. \label{fig:fig2}}
\end{figure}

The quality of the fits can be measured by reduced $\chi^2$ (which we
call $\chi_\nu^2$), the total $\chi^2$ divided by the number of
degrees of freedom.  The distribution of this quantity is shown in
Figure~\ref{fig:fig2}.  For the number of degrees of freedom
in our fitting scheme ($\sim 10^3$; the actual number varies over a
range of a factor of 3 for different objects), fits would ordinarily
be deemed acceptable only if $\chi_\nu^2$ is at most slightly greater
than unity.  None of our fits meets this test.

On the other hand, 90\% yield $\chi_\nu^2$ values $\leq 2.5$.  For
several reasons, we consider fits this good to be acceptable {\em for
our purposes}.  First of all, we are not claiming physicality for the
parameters (e.g., extinction) that we infer; all we really need is a
reasonable description of the flux before absorption in the vicinity
of the \ion{C}{4} and \ion{Mg}{2} lines.  Second, we are not computing
$\chi^2$ in strictly the standard sense since we weight some regions
of the spectrum differently than others.  Third, we are deliberately
not modeling certain features that can add to $\chi^2$, such as
intervening Ly$\alpha$ absorption and, of course, BAL absorption.
Fourth, we are compelled to make use of a wavelength band near
1300~\AA\ which we know to be contaminated by a variable amount of
emission line flux that we do not model.  Including this band is
necessary because fixing the global continuum shape across the
\ion{C}{4} region requires fitting to some region blueward of that
feature, and this band is the {\em least} contaminated band available.

When $\chi_\nu^2 > 2.5$, we visually inspected the spectra in order to
attempt an improvement.  If we believed that the fit in the \ion{C}{4}
region was reasonable (no matter what departures occurred elsewhere),
we accepted the fit, otherwise a new fit with zero extinction was
constructed.  If that resulted in an improved description of the
\ion{C}{4} region, we accepted the new fit.  If it did not, we marked
that object as having an unmeasurable balnicity.  Interestingly, every
quasar falling into this last category turned out to be an FeLoBAL.

In Figure~\ref{fig:fig2} we show the distribution of
$\chi^2_{\nu}$ for all of the quasars in the redshift range that we
have searched for BAL troughs (solid line).  The dashed line gives the
distribution for BALQSOs.  Note that the distribution for BALQSOs is
skewed towards slightly larger values --- as expected; absorption
features, by definition, will not match the continuum as defined by
the composite spectra and will contribute to an increase in $\chi^2$.
However, the similarity of the two distributions shows that BALs by
themselves are not the major contributor to $\chi^2$.

\subsection{Degeneracy of Spectral Index and Reddening\label{sec:degeneracy}}

\begin{figure}[p]
\epsscale{1.0}
\plotone{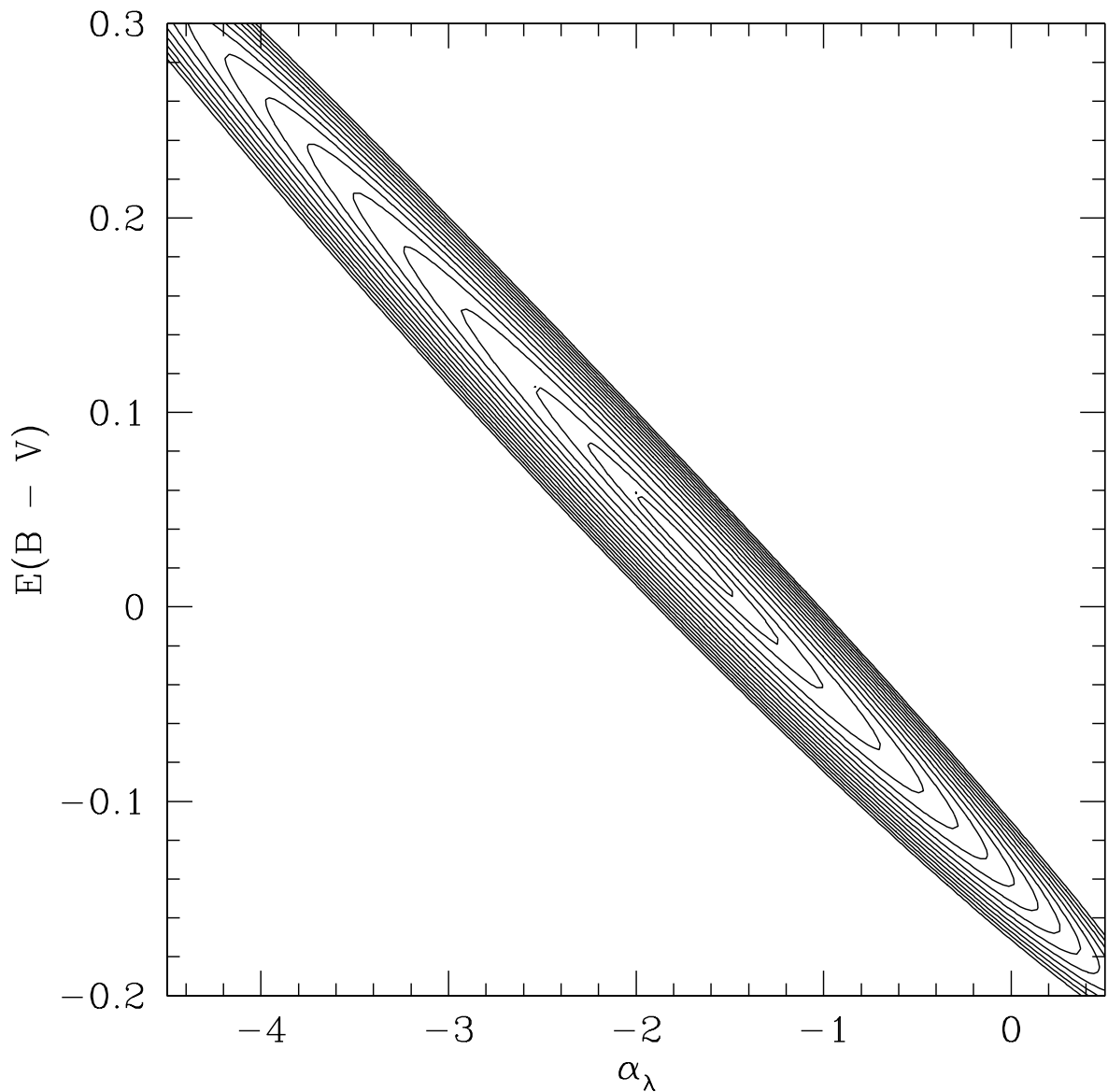}
\caption{A contour plot of a the reduced $\chi^2$ values that result
from fitting a composite spectrum continuum onto an individual
spectrum (SDSS J003019.82$-$002802.6, $\chi^2_{\nu} = 1.105$, top
panel of Fig.~\ref{fig:fig1}) by changing the spectral index and
reddening according to the SMC reddening law.  The contour levels
range from $\chi^2_{\nu} = 1.12$ (center curve) to $\chi^2_{\nu} =
2.5$ (outer curve). \label{fig:fig3}}
\end{figure}

It is perhaps not surprising that several combinations of the two fit
parameters (spectral index and reddening) produce satisfactory fits.
This result originates from the nature of the $\chi^2 (\alpha, E)$
function.  Figure~\ref{fig:fig3} shows a typical contour plot of the
paraboloid-like surface generated by the reduced $\chi^2$
distribution.  The surface is quite smooth and has a clear valley
along a line $\Delta \alpha = \beta E$, where
$\Delta\alpha \equiv \alpha - \alpha_0$ and $\beta \sim -10$ is the
tradeoff between
spectral index and reddening.\footnote{We use spectral indices in
$\lambda$-space.  In $\nu$-space,
$\Delta \alpha_\nu = -\Delta \alpha_\lambda$, and $\beta$ will change
sign accordingly.}  This line
has a small positive concavity and hence a local surface minimum at
the minimizing spectral index and reddening.  Although in principle
this local minimum represents the best-fit parameters, the small
curvature of the valley line allows nearby points on this line to
function, as well as the local minimum, as best-fit parameters.  More
specifically, small changes in the fit parameters from the local
minimum leave the reduced $\chi^2$ value nearly unchanged.

Deviations on the order of $\delta \alpha = \pm0.10$ and $\delta E =
\mp 0.010 \;(= \delta \alpha/\beta)$ produce a typical change in
reduced $\chi^2$ of $< 0.01$.  For example, for the quasar SDSS
J003019.82$-$002602.6, the minimum was found near $(\alpha, E,
\chi^2_{\nu}) = (-1.99, 0.055, 1.105)$, but the nearby point $(-1.90,
0.046, 1.107)$ is reasonably as good a fit (the corresponding
probability changes from 0.860 to 0.865).

\begin{figure}[p]
\epsscale{1.0} 
\plotone{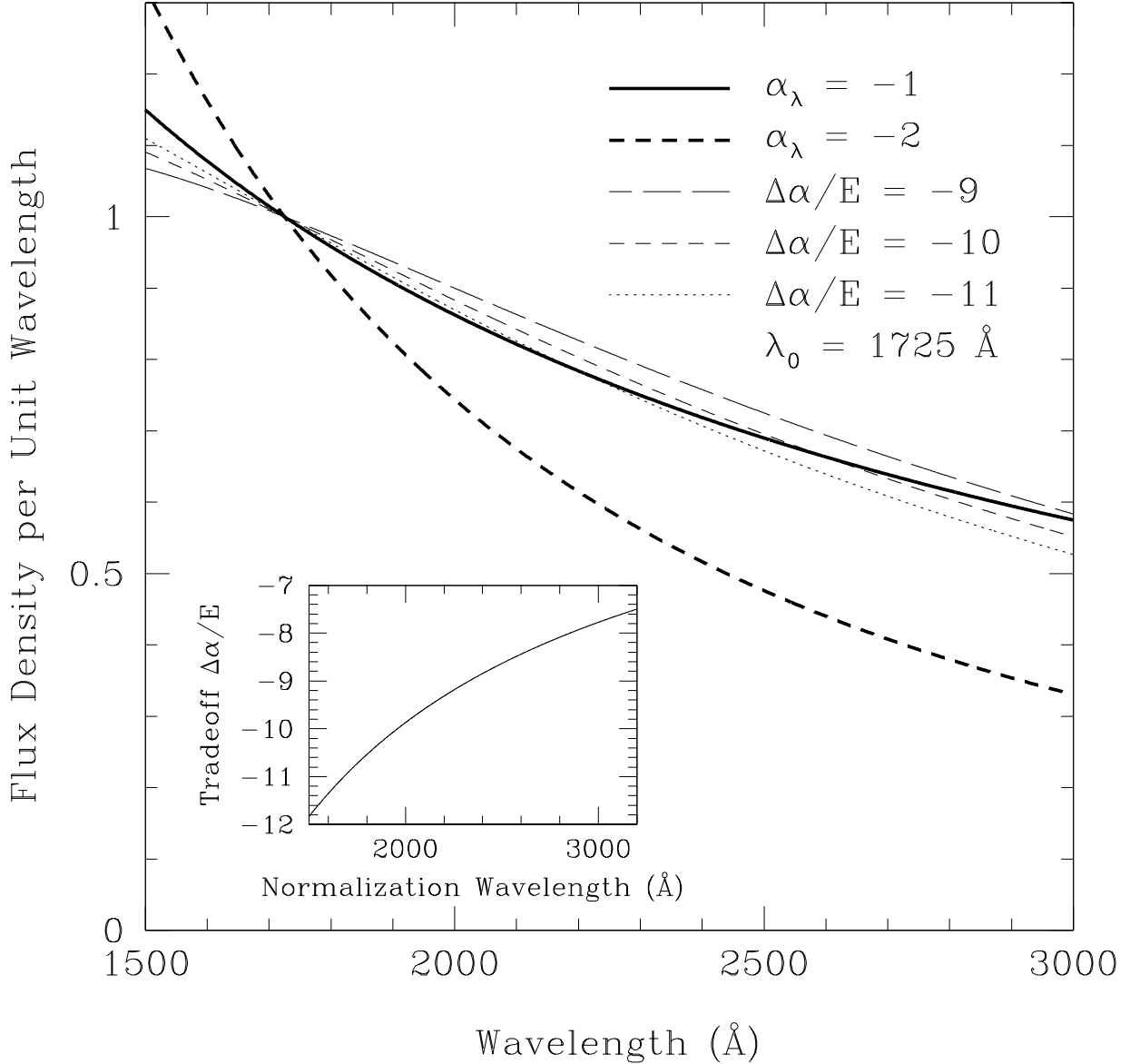}
\caption{The degeneracy of spectral index and reddening allows a blue
power law ($\alpha = -2$, {\em thick dashed line}) to approximate a
red power law ($\alpha = -1$, {\em thick solid line}) by reddening
according to an SMC-like reddening law.  The blue power law is
reddened with three amounts of reddening: $E(B - V)$ = 0.111 ({\em
thin long-dashed line}), 0.100 ({\em thin short-dashed line}), and
0.091 ({\em thin dotted line}), corresponding to $\beta = \Delta
\alpha/E = -9$, $-10$, and $-11$. ({\em Inset}) The mean tradeoff
$\beta = \Delta\alpha/E$ between spectral index and reddening for the
wavelength range $1500$ \AAA $< \lambda < 3200$ \AA.  A change in
spectral index can be compensated by a change in reddening multiplied
by the tradeoff factor, which depends on the normalization wavelength.
\label{fig:fig4}}
\end{figure}

The tradeoff between spectral index and reddening results from a
combination of the wavelength range included in the $\chi^2$ function
(where $w(\lambda) > 0$) and the normalization wavelength.  This can
be shown by equating a power law with spectral index $\alpha_0$ to an
SMC-reddened power law with spectral index $\alpha$ and reddening $E$
and solving for the tradeoff.  Upon normalizing both the power law and
reddened power law to unity at $\lambda_0$, the tradeoff (using a
power-law approximation to the SMC reddening law with exponent,
$D=1.2$ and $a=1.39$) is
\begin{equation}
\frac{\Delta \alpha}{E} =
\left(\frac{1}{\lambda^D}-\frac{1}{\lambda_0^D}\right)\frac{a}{\log_{10}
(\lambda/\lambda_0)}.
\end{equation}
We can take the arithmetic mean (via integration) of the tradeoff over
a wavelength range $\lambda_1 < \lambda < \lambda_2$ that includes the
\ion{C}{4} and \ion{Mg}{2} emission lines.  This yields the average
tradeoff
\begin{equation} 
\beta \equiv (\lambda_2-\lambda_1)^{-1} \int_{\lambda_1}^{\lambda_2} \frac{\Delta \alpha}{E} \; d\lambda.
\end{equation}
The average tradeoff remains a function of the normalization
wavelength $\lambda_0$.  Integrating over $1500$ \AAA $< \lambda <
3200$ \AA, $\beta$ is nearly linear in the same range of $\lambda_0$,
as shown in the inset of Figure~\ref{fig:fig4}.  We use three
normalization points at 1725 \AA, 2200 \AA, and 3150 \AA, and the
corresponding tradeoffs are $-10.8$, $-9.3$, and $-7.6$, respectively.
Thus when the composite-fitting algorithm fails to converge to the
best-fitting spectral index and reddening values, a simple power law
with spectral index $\alpha_0$ can be fitted to the curve. Then a new
initial guess of spectral index $\alpha$ and reddening $E = (\alpha -
\alpha_0)/\beta$ can be tried to promote convergence.

Figure~\ref{fig:fig4} illustrates how changing the the spectral
index of a power law and the reddening can mimic the original power
law.  We have plotted a pure power law with $\alpha = -1$ ({\em thick
solid line}).  To recover this red power law by using a blue power law
with $\alpha = -2$ ({\em thick dashed line}), the blue power law is
reddened by an amount determined by the tradeoff.  SMC-reddened power
laws with $E(B - V) = 0.091, 0.100$, and $0.111$, corresponding to
$\beta = \Delta \alpha/E = -11, -10,$ and $-9$, are over-plotted in
thin long-dashed, short-dashed, and dotted lines, respectively.  All
of the curves are normalized to unity at 1725 \AA.

In sum, although we regard our fits as providing a reasonable
description of the spectral shape, the best-fit parameters should not
be taken as physical --- these are {\it not} genuine measurements of
extinction; for this reason, we do not list them in our catalog.
However, even though the absolute values are not physical, the relative
values may still be meaningful; this possibility will be explored in
\markcite{rrh+02}{Reichard et al.} (2003) when we study the continuum properties of BALQSOs in
more detail.

\subsection{Emission Line Peak Adjustment\label{sec:empeak}}

Although the average quasar spectrum is representative of the whole
EDR quasar sample, in addition to differences in the slope and shape
of the continuum between individual quasars, there are also
differences in the emission line regions.  Since we are using the
average quasar spectrum as our template continuum+emission spectrum,
we must adjust this template in the emission line regions when the
strength of the emission lines in the template does not match the
strength in an individual quasar.  Fortunately, the equivalent width
distribution of broad lines in quasars spans only a factor of
$\sim10$, whereas the luminosity of quasars spans a few orders of
magnitude.

To improve the fitted composite spectrum as a continuum for balnicity
index measurement, we scale the \ion{C}{4} and \ion{Mg}{2} emission
lines to match the peak flux density of the input spectrum.  The
process is straightforward: after the composite is adjusted by the
fitting process described in \S~\ref{sec:fitcomposite}, the
normalized, reddened power law that is the result of that fitting is
subtracted from both the adjusted composite and the input spectra; we
then multiply the composite flux density in the emission line region
by the ratio of the peak flux densities in the line; and re-add the
normalized, reddened power law.  This method has the virtue of
automatically shifting the \ion{C}{4} line center relative to the
other emission line centers and also giving its profile the mean
asymmetry since the SDSS determines redshifts using empirical instead
of laboratory wavelengths for emission lines.  For example, in the
mean \ion{C}{4} is shifted $\simeq 800$~km~s$^{-1}$ blueward of
\ion{Mg}{2} \markcite{ric+02b}({Richards} {et~al.} 2002b) and $340$~km~s$^{-1}$ blueward of
\ion{C}{3}] \markcite{vrb+01}({Vanden Berk} {et~al.} 2001).

  Although this process yields line profiles that appear to be in
reasonably good agreement with the data, this approach is not perfect
and one can imagine better ways of adjusting the emission line
strengths.  In, particular, it is known that the strength of the
\ion{C}{4} emission line is dependent upon the luminosity of the
quasar \markcite{bal77,opg94}({Baldwin} 1977; {Osmer}, {Porter}, \& {Green} 1994) and the velocity offset of the \ion{C}{4}
peak with respect to \ion{Mg}{2} \markcite{ric+02b}({Richards} {et~al.} 2002b).  In addition, in the
composite method we make no allowance for any variation in either the
width of the \ion{C}{4} line or its shift relative to line-center as
found in the composite, even though both can vary considerably from
quasar to quasar \markcite{wil86}(e.g., {Wilkes} 1986).  Note that more conventional
methods \markcite{kro+02}(e.g., {Tolea} {et~al.} 2002) fit the width of the \ion{C}{4} line
but assume its line-center is at the same redshift as some standard
(e.g., \ion{C}{3}]1909), which may be shifted from systemic.  In
the future, we hope to account for emission line differences by having
not just one, but many template spectra.

\section{BALQSO Catalog\label{sec:catalog}}

\subsection{Balnicity Determination and Sample Definitions}

Once an intrinsic spectral shape has been chosen by our fitted
composite spectrum method, the balnicity of the input spectrum can be
determined.  Balnicity indices are computed for two lines: \ion{C}{4}
and \ion{Mg}{2}.  The algorithm was applied to each EDR quasar
spectrum, measuring the properties of any absorption troughs, i.e.,
determining if each spectrum is a HiBAL (broad absorption trough just
blueward of \ion{C}{4} emission), a LoBAL (broad absorption troughs
just blueward of both the \ion{C}{4} and \ion{Mg}{2} emission lines
--- but other lines such as \ion{Al}{3} could be used as well), or a
nonBAL (no broad absorption troughs just blueward of the \ion{C}{4}
and \ion{Mg}{2} emission lines).  The \ion{C}{4} balnicity index uses
the traditional \markcite{wmf+91}({Weymann} {et~al.} 1991) definition, and the \ion{Mg}{2}
balnicity index is a modified version (see below).  \markcite{hal+02}{Hall} {et~al.} (2002)
discuss how the determination of balnicity indices might be improved
in the future.
 
The fixed observed wavelength coverage of the SDSS EDR spectra
($3800\,{\rm \AA} \leq \lambda \leq 9200\,{\rm \AA}$) imposes
constraints on the redshifts at which BALQSOs can be identified and at
which balnicity indices can be calculated in our automated fashion.
First, for this catalog we restrict ourselves to identifying as
BALQSOs those quasars with \ion{C}{4} absorption at least $2000\,{\rm
km\,s^{-1}}$ broad located between $3000$ and $25,000\,{\rm
km\,s^{-1}}$ blueward of the quasar redshift (or \ion{Mg}{2}
absorption $1000\,{\rm km\,s^{-1}}$ broad between $0$ and
$25,000\,{\rm km\,s^{-1}}$ blueward of the quasar redshift; see
below).  Second, complete samples of BALQSOs can be identified only
for redshifts at which the EDR spectra contain the entire
$25,000\,{\rm km\,s^{-1}}$ range which is searched for BAL
troughs.\footnote{These samples could still be slightly incomplete, as
BAL troughs are known to exist at higher outflow velocities.}  Third,
for our automated balnicity index calculation, the redshifts must be
restricted so that the normalization windows at $1725\,{\rm \AA}$ (for
\ion{C}{4}) and $3150\,{\rm \AA}$ (for \ion{Mg}{2}) are observed.
These constraints restrict the redshift range to $1.7 \leq z \leq 4.2$
for \ion{C}{4} and $0.5 \leq z \leq 1.9$ for \ion{Mg}{2}.  We can
extend this latter range to $0.5 \leq z \leq 2.1$ by normalizing at
$2200\,{\rm \AA}$ for $z > 1.9$.  BALQSOs with redshifts outside of
these ranges can also be identified, but not in a complete sense.

We modified our \ion{Mg}{2} balnicity index definition from the
\ion{C}{4} definition in two ways.  \ion{Mg}{2} broad absorption
troughs tend to be weaker and narrower than those found in \ion{C}{4}
troughs \markcite{vwk93}({Voit}, {Weymann}, \& {Korista} 1993).  The minimum \ion{Mg}{2} continuous absorption
width was chosen to be 1000 km~s$^{-1}$.  Moreover, the \ion{Mg}{2}
absorption removes flux density at smaller velocity displacements (on
average) from the center of the emission line than is the case in
\ion{C}{4}.  We accordingly adjust the minimum velocity from 3000
km~s$^{-1}$ to 0 to include low-velocity absorption in the \ion{Mg}{2}
balnicity index (see \markcite{hal+02}{Hall} {et~al.} 2002 for further justification of
these modifications).  The \ion{Mg}{2} balnicity indices were computed
in the same manner as for \ion{C}{4} balnicity indices but with
normalization at 3150 $\rm{\AA}$ or 2200 $\rm{\AA}$ (depending on the
quasar redshift).

Since all known quasars that exhibit low-ionization BAL troughs also
exhibit high-ionization BAL troughs, we limited our \ion{Mg}{2} BALQSO
sample to objects that we have identified as \ion{C}{4} BALQSOs and
those objects with redshifts $0.5 \le z \le 1.7$ whose \ion{C}{4}
balnicity index cannot be computed.  Our HiBAL sample will be broken
into subsamples including and excluding LoBALs.  Note that our
semi-automated algorithm does not make an attempt to classify objects
as LoBALs based on \ion{Al}{3} absorption.  Since \ion{Al}{3}
absorption can be stronger than \ion{Mg}{2} absorption in LoBALs, our
method will miss some fraction of the LoBALs.  However, our by-eye
classification should have recovered most of those \ion{Al}{3} LoBALs
missed by our semi-automated algorithm.

\subsection{Construction of the Catalog}

\ion{C}{4} BALQSOs were selected from the EDR sample first by choosing
those objects with a non-zero balnicity index as determined by either
the composite method, the method of \markcite{kro+02}{Tolea} {et~al.} (2002), or visual
inspection by one of us (PBH).  The two automated methods selected
very nearly the same objects over their common redshift range.
Spectra of each of these BALQSO candidates were then inspected by eye.

A small number of objects were discarded from the initial
composite-generated BALQSO list because the composite spectrum
continuum poorly matched the object's continuum, and the balnicity
calculation was suspect.  The poor quality of the fit of the composite
was typically the result of one of three causes.  First, a few objects
had extraneous emission or absorption at the normalization wavelength
that shifted the fitted composite spectrum above or below the object
spectra.  NonBALQSOs with a high normalization will appear to have
nonzero balnicity indices.  Second, the $\chi^2$-minimization and
$(\alpha, E)$ point-scanning algorithms failed on some spectra that
either had sufficiently strong absorption that a power law is not
descriptive of the continuum, e.g., FeLoBALs, or the algorithms simply
did not find a minimizing pair of fit parameters.  These objects were
reclassified by inspection as BALQSOs or nonBALQSOs.  Third, a few
objects with $z\sim1.7$ had excellent composite fits redward of the
\ion{C}{4} emission line but with an overestimated continuum in the
\ion{C}{4} broad absorption region, mainly because little of the
weighted wavelength ranges were observed in these spectra blueward of
the broad absorption region.  We replaced these composite fits (fitted
with the combination SMC reddening law and power law) with a composite
fitted using a power law without reddening, recomputed the balnicity
index, and reclassified the object.  These three errors were more
prominent in computing \ion{Mg}{2} balnicity index, so we classified
quasars as LoBALs only after visual inspection rather than relying
solely on measured \ion{Mg}{2} balnicity indices.

We present a complete catalog of BALQSOs from the EDR in the redshift
range $1.7 \le z \le 4.2$ in Table~\ref{tab:tab1}.  The BAL catalog
consists of 185 BALQSOs, including 153 HiBALs, 24 LoBALs, and 8
FeLoBALs.  Note that we include six quasars in Table~1 that formally
have BI=0 (or where the BI is not measured).  Three of them are LoBALs
or FeLoBALs and should clearly be included.  The three HiBALs should
be excluded if a pure sample defined strictly using the \markcite{wmf+91}{Weymann} {et~al.} (1991)
BI criteria is desired, but we believe that they should be included
because they show troughs that came very near to satisfying the
balnicity criteria.  For example, there is a clear C~IV absorption
trough in SDSS~110838.76$-$005533.7, but it extends to a velocity
larger than $v=25,000\,{\rm km\,s^{-1}}$, which throws off the
automated fit.

In Table 1, column 1 lists the object names using the SDSS format of
J2000 right ascension (hhmmss.ss) and declination ($\pm$ddmmss.s).
Columns 2$-$4 list the plate, fiber, and modified Julian date.
Columns 5$-$7 indicate whether an object was identified as a quasar
candidate by the final version of the quasar target selection
algorithm (Column 5, \markcite{rfn+02}{Richards} {et~al.} 2002a), by
the EDR quasar target selection algorithm (Column 6), and by the EDR
serendipity target selection algorithm (Column 7).  Of the 185 BALQSOs
in Table~1, Column~5 indicates that 131 of them meet the SDSS's
adopted quasar target selection criteria and thus constitute a more
homogeneous subsample.  Column 8 gives the FIRST peak flux density at
20 cm; 16 of our BALQSOs are radio-detected, only three of which
overlap with \markcite{mvi+01}{Menou} {et~al.} (2001).  A zero in this
column means that the object is undetected in the FIRST survey at the
$\sim1\,{\rm mJy}$ limit; no data indicates that there is no FIRST
observation at this location.  Columns 9, 10 and 11 list the redshift,
the apparent $i^*$ magnitude (corrected for Galactic reddening;
\markcite{sfd98}{Schlegel}, {Finkbeiner}, \& {Davis} 1998) and the
resulting absolute $i^*$ magnitude, respectively.  Columns 12$-$14
list the balnicity indices as follows: Columns 12 and 14 list
\ion{C}{4} and \ion{Mg}{2} balnicity indices computed by the
composite-fitting algorithm (this paper), while Column 13 gives the
\ion{C}{4} balnicity index as computed by using the traditional method
by \markcite{kro+02}{Tolea} {et~al.} (2002) for objects with $1.8 \le
z \le 3.8$.

The subsequent four columns show the classification of the objects.
Column 15 gives the classification according to our fitted composite
algorithm: H = HiBAL and N = nonBAL.  Column 16 shows the
classification according to \markcite{kro+02}{Tolea} {et~al.} (2002).  All of the objects are
labeled in this column as either ``H'' or ``N'' with similar meanings
to those in Column 15.  The two automated algorithms define continua
in different ways.  Because the continua are not identical, there are
cases where they disagree on whether an object is a BALQSO.  In the
catalog we include all quasars defined as BALQSOs by either algorithm.

One of us (PBH) examined all of the EDR spectra for BALQSOs by eye;
the resulting classifications are listed in Column 17.  The following
abbreviations are used: ``HiZ'' = HiBAL at $z\ge3.90$, where LoBALs
could be missed because $\lambda_{\rm AlIII} > 9150$ \AA; ``Hiz/Loz''
= Hi/LoBAL at $z\ge2.26$, where LoBALs could be missed because
$\lambda_{\rm MgII} > 9150$ \AA; ``Hi'' = only high-ionization lines
are present; ``Lo'' = both high- and low-ionization lines are present;
``LoZ'' = LoBAL at $z\ge3.90$, where LoBALs could be missed because
$\lambda_{\rm AlIII} > 9150$ \AA, ``FeLo'' = high-, low-, and
metastable \ion{Fe}{2}/\ion{Fe}{3} lines ($\sim 2600$ \AA, or
otherwise) or atypical absorption that more readily classifies the
object as an unusual or extreme BALQSO, usually with strong iron
absorption; and ``no'' = no broad absorption lines are present
(nonBAL).  A question mark ``?'' means uncertainty in the previous
code; e.g., ``HiLo?'' means a tentative LoBAL trough in a definite
HiBAL.

Finally, Column 18 gives the overall classification upon which we have
decided for each quasar.  It lists an object as a LoBAL (``Lo'') if
visual inspection classified the object as a LoBAL.  Visual inspection
also revealed a few unusual LoBALs with iron absorption (FeLoBALs),
which are labeled instead by ``FeLo''.  The remaining objects are
those classified as HiBALs by at least one of the three classification
methods and are labeled ``Hi''.  We have supplied \ion{Mg}{2}
balnicity indices only for those objects classified as LoBALs in the
redshift range $0.5 \le z \le 2.1$

We also present a supplementary catalog of BALQSOs with redshift
generally outside the range $1.7 \le z \le 4.2$ in
Table~\ref{tab:tab2}, but with absorption clearly broad enough to
be considered as BALQSOs; the format is the same format as in
Table~\ref{tab:tab1}.  The supplement contains 39 BALQSOs,
including 27 HiBALs, 10 LoBALs, and 2 FeLoBALs --- all of which were
found by visual inspection.  One object in this supplement (SDSS
J235238.08+010552.4) has an intermediate redshift $z = 2.156$ and two
prominent \ion{C}{4} broad absorption troughs.  We have included this
object in our supplement rather than the main catalog because one
trough has a velocity shift too low, and the other too high, for the
object to qualify as a BALQSO by the traditional definition.  Two of
the radio-detected quasars in this supplementary sample are also
included in the \markcite{mvi+01}{Menou} {et~al.} (2001) analysis.  Between
Tables~\ref{tab:tab1} and~\ref{tab:tab2} we catalog 224
BALQSOs.

\subsection{NonBAL Sample \label{sec:nonbalsample}}

We wish to compare our classes of BALQSOs with a large class of
objects with no broad absorption lines.  To avoid rediscovering or
confusing correlations with luminosity and redshift when comparing
nonBALs and BALQSOs, we have created a nonBALQSO sample with
essentially identical distributions of absolute magnitudes $M_{i^*}$
and redshifts in the BALQSO sample.

For each BALQSO (HiBALs and LoBALs), we searched the sample of
nonBALQSOs and selected the object with the most similar absolute
$i^*$ magnitude and redshift without choosing any nonBALQSO more than
once.  We repeated this process three more times with the remaining
nonBALQSOs (without duplication).  Each object in the BALQSO sample
was thus matched to four nonBALQSOs with similar absolute magnitude
and redshift, yielding a nonBALQSO sample with four times as many
objects as the BALQSO sample.

Typical deviations in $M_{i^*}$ and $z$ between a pair of matched BAL
and nonBALQSOs are $\Delta M_{i^*} \sim 0.06$ mag (slightly higher
than 0.03, the systematic photometric error) and $\Delta z \sim 0.06$.
We performed K-S and Student-$t$ tests \markcite{ptv+92}({Press} {et~al.} 1992) to ensure that
the magnitudes and redshifts of the BAL and nonBALQSO samples were
consistent.

\subsection{BAL and nonBALQSO Composite Spectra}

\begin{figure}[p]
\epsscale{1.0} 
\plotone{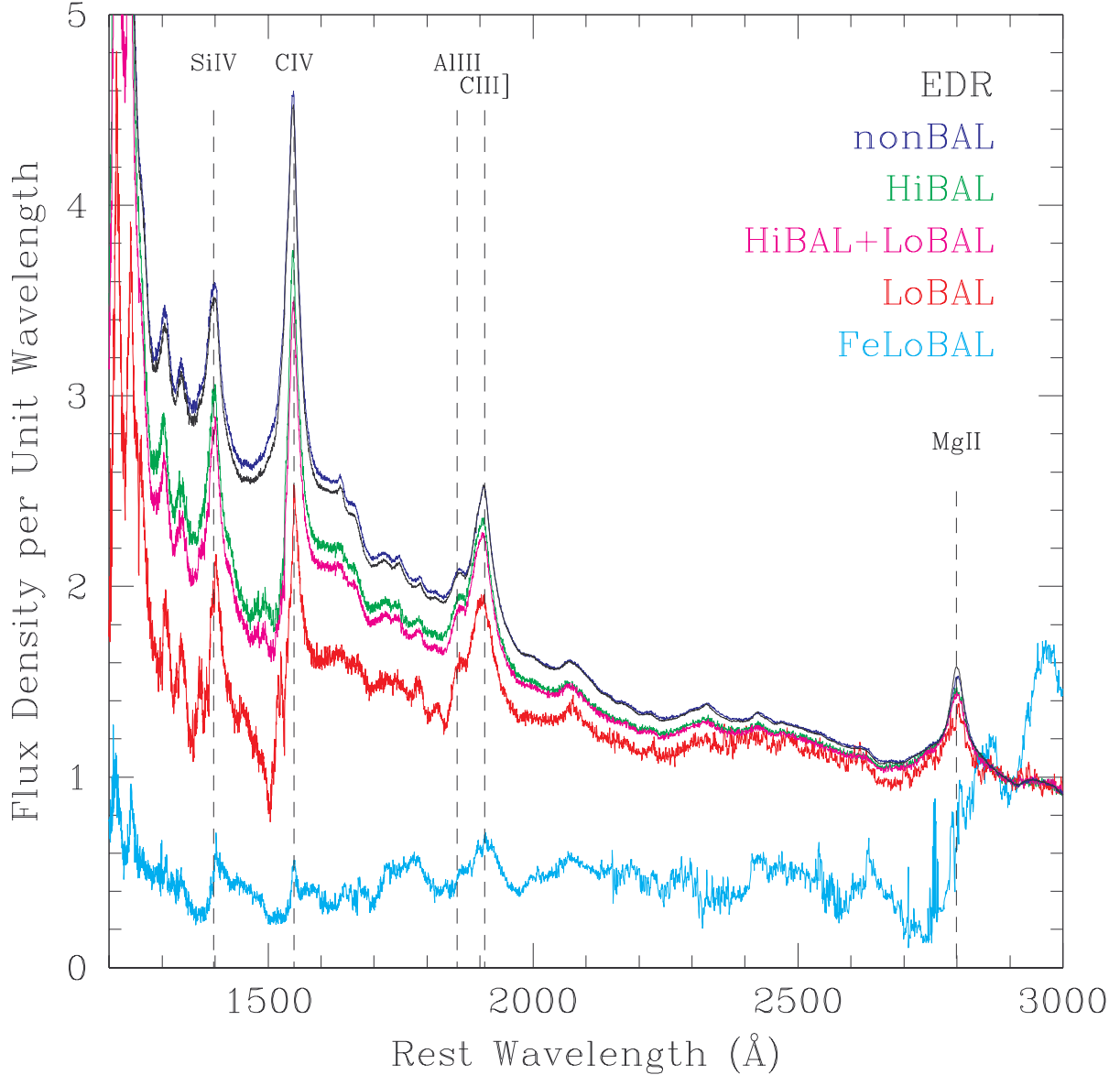}
\caption{The normalized composite spectra of the full EDR quasar
sample ({\em black}), nonBALs ({\em blue}), HiBALs (including LoBALs,
{\em magenta}), Hi\-BAL/non\-LoBALs (excluding LoBALs, {\em green}),
LoBALs ({\em red}), and FeLoBALs ({\em cyan}).  The spectra are
similar at long wavelengths above 2400 \AAA (with the exception of the
FeLoBAL composite --- which also shows emission near 2950 \AA), but
the BALQSO composite spectra show clear flux deficits at shorter
wavelengths as compared to the nonBALQSO composite spectrum; see
\markcite{hal+02}{Hall} {et~al.} (2002) for further details with regard to the absorption
structures seen in FeLoBALs.  Note that the redness of the BALQSO
spectra is not simply the result of the BAL absorption troughs, but
rather because the SEDs themselves are redder. \label{fig:fig5}}
\end{figure}

To compare the populations of nonBALQSOs, HiBALs, and LoBALs, we
created geometric mean composite spectra for each of these classes of
objects.  We use the method of \markcite{vrb+01}{Vanden Berk} {et~al.}
(2001), whereby all spectra from a sample are deredshifted,
normalized, binned into wavelength ranges, and geometrically averaged
together bin by bin.  The HiBAL composite was created from all objects
labeled by ``Hi'' in Column 18 of Tables~\ref{tab:tab1}
and~\ref{tab:tab2}, the LoBAL composite by those labeled ``Lo'', the
HiBAL+LoBAL composite by those marked ``Hi'' or ``Lo'', and the
FeLoBAL composite by those labeled ``FeLo''.  The nonBALQSO composite
spectrum was created from the matched sample of nonBALQSOs discussed
above.  These composite spectra are shown in Figure~\ref{fig:fig5}.
Note that the redness of the BALQSO spectra is not simply the result
of the BAL absorption troughs, but rather because the SEDs themselves
are redder.  If interpreted as extinction, for HiBALs $E(B-V) \sim
0.023$ and for LoBALs $E(B-V) \sim 0.077$ (assuming SMC-like dust
extinction), but more careful analysis is required to clarify what
this means; see \markcite{rrh+02}{Reichard et al.}  (2003), which will
present a comparison of the continuum and emission line features of
these composites.  In addition, Figure~\ref{fig:fig5} shows that the
\ion{C}{4} absorption troughs extend to within $3000\,{\rm
km\,s^{-1}}$ of the peak of the emission line (starting at essentially
$0$ velocity for the LoBAL composite); see below for further
discussion.

\section{Discussion \label{sec:discussion}}

\subsection{BI Comparison \label{sec:methodcomp}}

\begin{figure}[p]
\epsscale{1.0}
\plottwo{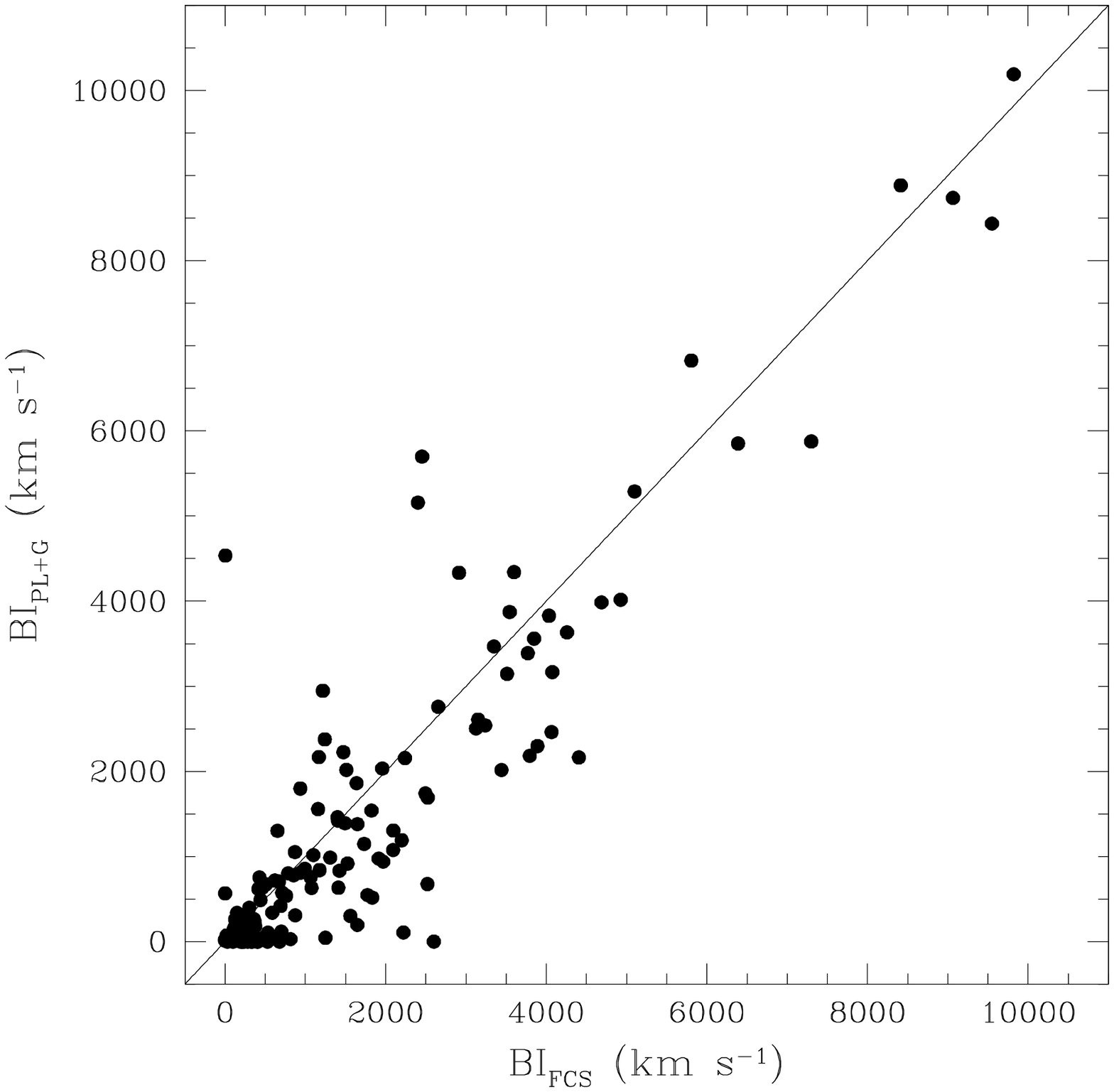}{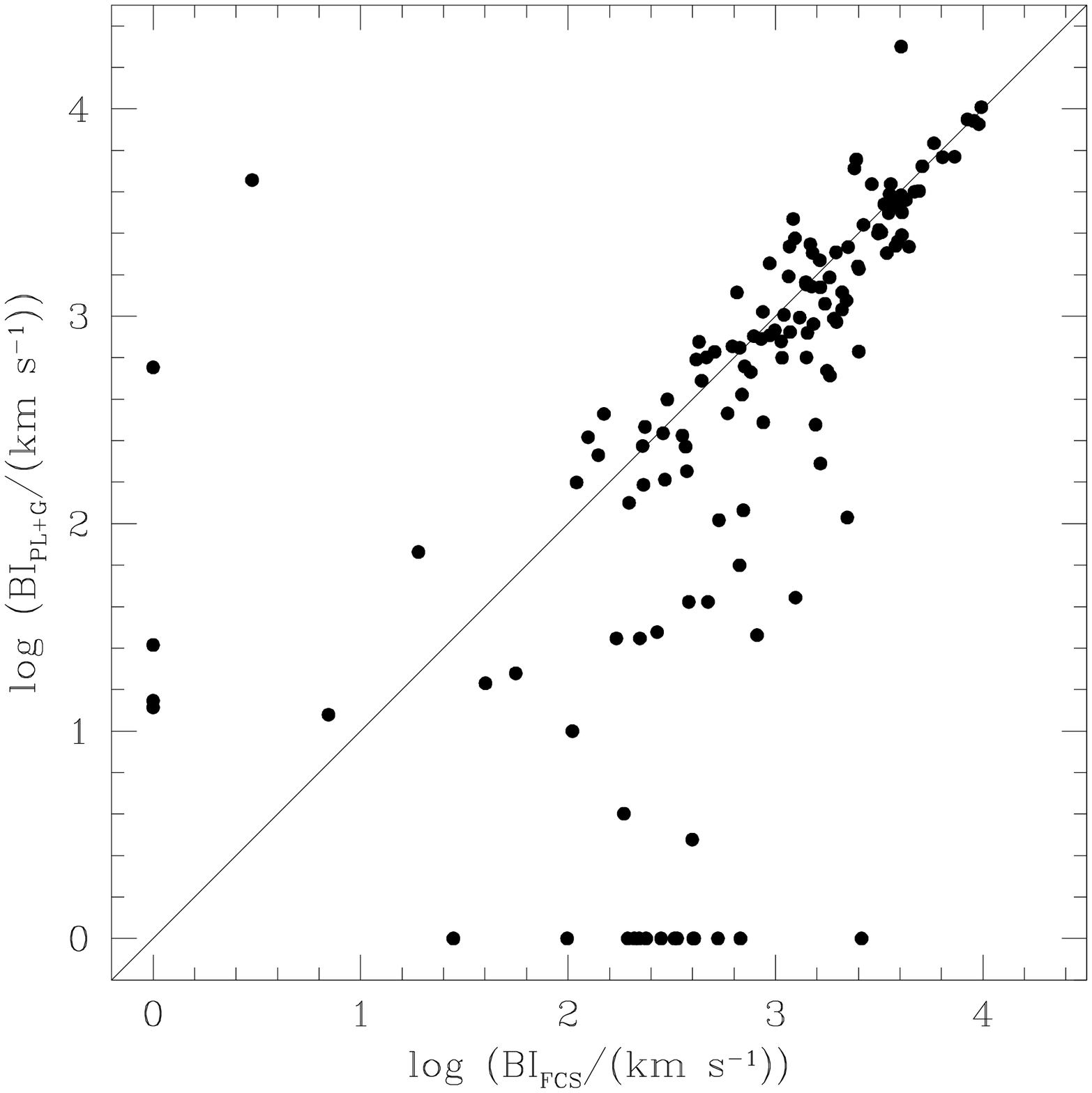}
\caption{Comparison of the \ion{C}{4} balnicity indices computed by
using a fitted composite spectrum ($BI_{FCS}$) and by using a power
law + Gaussian line ($BI_{PL+G}$). The left-hand panel shows the
balnicity indices in a linear representation; the right-hand panel
shows the same data in a log-log representation, which emphasizes the
small BI end of the distribution.  Quasars that were found to have
$BI=0$ by either method but that we believe to be true BALQSOs are
plotted with $\log(BI)=0$ on the appropriate axis.
\label{fig:fig6}}
\end{figure}

Before we analyze the BI distribution, we first compare our BI values
to those from \markcite{kro+02}{Tolea} {et~al.} (2002).  In Figure~\ref{fig:fig6} the \ion{C}{4}
balnicity indices computed using the fitted composite spectrum (FCS)
method from the current paper are plotted against the traditional
power law + Gaussian line (PL+G) results from \markcite{kro+02}{Tolea} {et~al.} (2002).  The rms
of the differences between the two samples is $945\,{\rm km\,s^{-1}}$,
but it is not uniform with BI: the rms fractional discrepancy
decreases from $\simeq 1$ for those with BI $< 300$~km~s$^{-1}$ to
$\simeq 0.25$ for the largest BIs measured ($ > 4,000$~km~s$^{-1}$).
Although the scatter about the line of unit ratio between the two BI
measures is nearly symmetric, there is a tendency for the FCS method
to give larger BIs to those BALs the PL+G method would assign a value
$<100$~km~s$^{-1}$.

For single trough HiBALs the scatter in the points from the $BI_{PL+G}
= BI_{FCS}$ line is due mainly to the difference in flux levels from
which absorption was measured.  BALQSOs with more discrepant BI values
can be attributed to a number of causes, most of which are generally
related to the placement of the continuum.  For all FeLoBALs and some
LoBALs, broad absorption can nearly eliminate the object's continuum
in parts of the spectrum.  In these cases, fitting a composite
spectrum or even a power law to the object is difficult, and there is
little chance that the fitted composite spectrum and fitted power law
will define similar continua through the \ion{C}{4} broad absorption
line.  Fortunately, this problem is mitigated to some extent by the
fact that any reasonable estimate of the continuum will yield a BI
that is appropriately large even if it is not entirely accurate.

In addition, objects with narrower absorption troughs with widths near
2000 \kmss (the lower limit on trough widths) may be given a zero
balnicity index by one method and a nonzero balnicity index by the
other.  These quasars might be classified as a nonBALQSO or BALQSO
depending on exactly how the flux level before absorption was
determined (there are 4 quasars for which the PL+G method finds
non-zero BI values and the FCS method finds $BI=0$, and 13 that are
the other way around).  BALQSOs with multiple or complex troughs also
present some problems.  Sometimes the continuum placement will be such
that one method finds that the entire absorption complex contributes
to the BI, whereas another continuum placement will count only part of
the absorption complex.  Therefore, one also expects that there will
be some large discrepancies for objects with relatively large BI
values.

Finally, it is important to realize that the redshifts used in the
composite method and the redshifts used by \markcite{kro+02}{Tolea} {et~al.} (2002) were derived
in slightly different ways.  In the mean, the \ion{C}{4} and
\ion{C}{3}] redshifts differ by $\simeq 340$~\kms, but in individual
objects the offset of \ion{C}{4} from systemic can be up to
$\sim2000\,{\rm km\,s^{-1}}$ \markcite{ric+02b}({Richards} {et~al.} 2002b).  In the FCS method the
redshifts are taken from the EDR quasar catalog \markcite{sch+02}({Schneider} {et~al.} 2002) and
were determined by using the {\em empirical} wavelengths of the
emission lines commonly seen in quasars \markcite{vrb+01}({Vanden Berk} {et~al.} 2001) as opposed to
the laboratory wavelengths --- thus partially correcting for emission
line shifts.  On the other hand, the PL+G method directly fit to the
emission line profiles of each individual quasar, rather than assuming
that all shared the velocity width found in the composite.  These
different treatments of the emission line profile are especially
important, of course, to BALQSOs with small BIs.  It is likely that
the relatively large discrepancies found for small BI cases are due to
these systematic contrasts.

In conclusion, because the two automated methods disagreed about the
{\it existence} of a BAL feature in only very few cases, we believe
that they have mutually validated each other as methods for deciding
whether broad absorption is present in quasar spectra.  In addition,
when the feature is strong, the two methods agree reasonably well with
regard to its magnitude (the BI measure).  However, for weaker
features, there can be significant discrepancies between the BIs
produced by the two methods; we must therefore acknowledge a
significant systematic uncertainty in the smaller BI range.  It is
exactly for these cases that the arbitrariness in the parameters of
the classical BI definition (e.g., the minimum offset, the minimum
width of continuous absorption) creates at least as great a systematic
uncertainty in determining the {\it physical} ``balnicity"
\markcite{hal+02}({Hall} {et~al.} 2002).

\subsection{Balnicity Index Distribution \label{sec:bidist}}

\begin{figure}[p]
\epsscale{1.0} 
\plotone{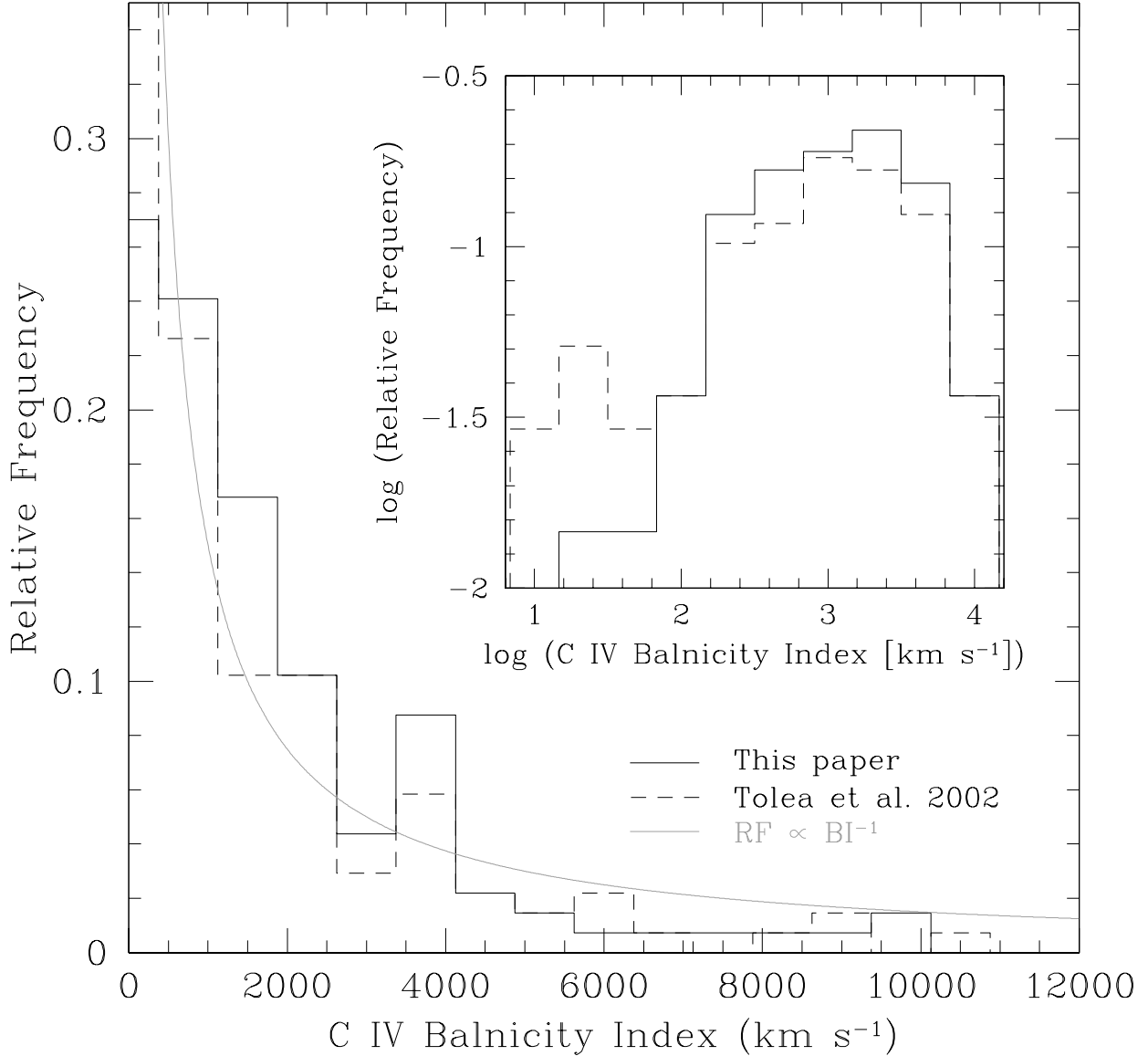}
\caption{The C IV balnicity index distribution of BALQSOs.  These
histograms show the distribution of the balnicity indices as computed
with the fitted EDR composite spectrum ({\em solid histogram}) and
using a power law + Gaussian line ({\em dashed histogram},
\markcite{kro+02}{Tolea} {et~al.} 2002).  The solid gray line depicts a distribution that is
proportional to BI$^{-1}$; it is not a fit to the data, but rather it
is merely meant to guide the eye.  The inset is a log-log version of
the plot; it emphasizes the smaller end of the BI distribution and
shows an apparent change in BI distribution towards smaller BI
($\lesssim1500$ \kms; not to be confused with the differences between
the two methods for BI $\lesssim100$ \kms).\label{fig:fig7}}
\end{figure}

The distribution of balnicity indices using the FCS (solid histogram)
and PL+G (as measured by \markcite{kro+02}{Tolea} {et~al.} 2002, dashed histogram) methods
to define continua are shown in Figure~\ref{fig:fig7}.  The two
distributions are very similar; each shows a large fraction of BALQSOs
with small balnicity indices and a shallow tail of BALQSOs with large
balncities.  The BI distribution goes roughly as $BI^{-1}$, where the
$-1$ power-law index (gray line in Fig.~\ref{fig:fig7}) is {\em not} a
fit to the data, but rather is meant to guide the eye.  The inset to
Figure~\ref{fig:fig7} shows that both the distribution appears to
change significantly for moderate BI ($\lesssim1500$ \kms) and that
the PL+G method yields more objects with BI $\lesssim100$ \kms; see
below for more details.

Comparison of these distributions with that from the \markcite{wmf+91}{Weymann} {et~al.} (1991)
sample show significant differences.  \markcite{wmf+91}{Weymann} {et~al.} (1991) computed
\ion{C}{4} balnicity indices for 42 BALs.  Two BALs (5\%) were
``borderline'' cases, with \ion{C}{4} balnicity indices $< 600$ \kms,
18 BALs (43\%) had balnicity indices $> 5000$ \kms, and three BALs
(7\%) had balnicity indices $> 10,000$ \kms.  In our sample, we find
that \ion{C}{4} balnicity indices are generally smaller.  Only two
objects (1\%) out of the 185 in our complete BAL sample
(Table~\ref{tab:tab1}) possessed a \ion{C}{4} balnicity index above
10,000 \kmss (these were further classified by eye as LoBALs), and only
14 (8\%) have a balnicity index above 5000 \kms.  The differences
between the balnicity index distribution of \markcite{wmf+91}{Weymann} {et~al.} (1991) and ours
could be the result of the differences in continuum and emission line
fitting methods, small number statistics, or our sample being more
complete to quasars with lower BI values.

The \ion{C}{4} balnicity indices of our LoBAL sample of 24 objects
(from Table~\ref{tab:tab1}) were also generally less than those of
the entire BALQSO sample of \markcite{wmf+91}{Weymann} {et~al.} (1991).  Our LoBAL sample has 6
objects (25\%) with \ion{C}{4} balnicity indices $> 5000$ \kms, 2 of
the latter having \ion{C}{4} balnicity indices over 10000 \kms.  We
find that the median LoBAL \ion{C}{4} balnicity index is 3544 \kmss as
computed using the fitted composite spectrum continuum and 3468 \kmss
using the power law + Gaussian line continuum; the HiBAL medians are
937 \kmss and 567 \kms, respectively.  LoBALs exhibit much deeper
\ion{C}{4} broad absorption troughs than Hi\-BAL/non\-LoBALs, as noted
by \markcite{wmf+91}{Weymann} {et~al.} (1991), which is normally interpreted as evidence that
HiBALs and LoBALs are not distinct types, but rather represent the
extremes of a continuum in absorption strength/ionization.

The fact that much of our sample has rather small BI strongly suggests
a connection between BAL troughs and the so-called ``associated
absorbers'' \markcite{fwp+86}({Foltz} {et~al.} 1986) that occur within $\sim3000\,{\rm
km\,s^{-1}}$ of the emission redshift of some quasars (particularly
steep-spectrum radio-loud quasars).  However, we stress that the
details of the shape of the BI distribution for $BI \ll
1000$~km~s$^{-1}$ are not well-determined.  Part of the uncertainty is
the relatively large systematic error discussed in
\S~\ref{sec:methodcomp}.  A larger problem lurks in the arbitrariness
of the BI definition.  If the minimum width of continuous absorption
were fixed at, say, 1500~km~s$^{-1}$, numerous additional quasars with
weaker absorption (or choppier absorption profiles) would be
classified as BALQSOs.  In fact, the large number of small BI BALQSOs
means that, in this work and all previous work that imposed the
\markcite{wmf+91}{Weymann} {et~al.} (1991) balnicity criteria, the fraction of quasars with
intrinsic outflows has been {\em significantly} underestimated since
those quasars that just barely fail to satisfy the BAL criteria would
populate the small BI end of the distribution where the density of
objects per BI is greatest.

Furthermore, the LoBAL composite spectrum and both HiBAL composite
spectra show significant broad absorption within 3000 \kmss of the
\ion{C}{4} emission line peak.  Thus, the balnicity index definition
given by \markcite{wmf+91}{Weymann} {et~al.} (1991) needs adjustment if we are to define an index
that includes {\em all} BAL-like outflowing absorption, including that
in SDSS~J235238.08+010552.4 of Table~\ref{tab:tab2} (which has
very high velocity absorption as opposed to very low velocity
absorption).  Suggestions for alternative balnicity indices were
proposed by \markcite{hal+02}{Hall} {et~al.} (2002) and will be considered in more detail in
future work on SDSS BALQSOs.

\section{Conclusions \label{sec:conclusions}}

We have presented a catalog of 224 BALQSOs from the SDSS Early Data
Release Quasar Catalog.  This sample represents a significant increase
in the total number of cataloged BALQSOs.  Of these 224 BALQSOs, 185
occupy a region in redshift space where our catalog should be
relatively complete to most BALQSOs.  Imposing further constraints
upon the initial selection of these objects as quasar candidates
leaves us with a relatively homogeneous catalog with 131 BALQSOs.

A comparison of the catalog that results from using our fitted
composite spectrum continuum method of classifying quasars as BALQSOs
to the BALQSOs sample constructed by \markcite{kro+02}{Tolea} {et~al.} (2002) using the more
traditional power-law plus Gaussian continuum method demonstrates that
the two methods agree reasonably well when the absorption is
relatively strong.  Given this agreement, extending our method to make
use of multiple templates should allow for even better and more
sensitive BALQSO classification in the future.

In agreement with \markcite{kro+02}{Tolea} {et~al.} (2002), we find that the distribution of
balnicity indices rises rapidly with decreasing balnicity index.  Our
\ion{C}{4} balnicity indices tend to be less than 5000 \kmss and hence
less than those given by \markcite{wmf+91}{Weymann} {et~al.} (1991).  Furthermore, even though we
ignored absorption within 3000 \kmss of the adopted redshift when
defining our BALQSO sample, \ion{C}{4} broad absorption troughs extend
within 3000 \kmss of the expected emission center for BALs, and the
absorption in LoBALs can start at velocities as low as 0 \kmss from
the expected emission center.  The combination of these results
strongly suggests that the \markcite{wmf+91}{Weymann} {et~al.} (1991) BALQSO definition is too
strict in terms of identifying quasars with intrinsic outflows and
that the true fraction of quasars with such outflows may be
significantly larger than estimated in the past (or herein).  If there
are numerous quasars with comparatively weak ``broad'' absorption, it
will, of course, be very difficult to separate them from those with
genuine ``associated absorbers''.  It is possible that the associated
absorber ($z_{em} \simeq z_{abs}$) population is merely the weak tail
of the BALQSO distribution (and that FeLoBALs and LoBALs are simply
the strong tail of the same distribution).

\acknowledgements

Funding for the creation and distribution of the SDSS Archive has been
provided by the Alfred P. Sloan Foundation, the Participating
Institutions, the National Aeronautics and Space Administration, the
National Science Foundation, the U.S. Department of Energy, the
Japanese Monbukagakusho, and the Max Planck Society. The SDSS Web site
is http://www.sdss.org/.  The SDSS is managed by the Astrophysical
Research Consortium (ARC) for the Participating Institutions. The
Participating Institutions are The University of Chicago, Fermilab,
the Institute for Advanced Study, the Japan Participation Group, The
Johns Hopkins University, Los Alamos National Laboratory, the
Max-Planck-Institute for Astronomy (MPIA), the Max-Planck-Institute
for Astrophysics (MPA), New Mexico State University, University of
Pittsburgh, Princeton University, the United States Naval Observatory,
and the University of Washington.  T.~A.~R., G.~T.~R., and
D.~P.~S. were supported by National Science Foundation grant
AST99-00703.  P.~B.~H. is supported by FONDECYT grant 1010981.  We
thank Niel Brandt for reading an early draft of the manuscript.

\clearpage



\clearpage

\clearpage

\clearpage

\clearpage
 
\clearpage

\clearpage

\clearpage

\clearpage

\begin{deluxetable}{lrrrcccrcccrrrcclc}
\rotate
\tabletypesize{\tiny}
\tablewidth{0pt}
\tablecaption{SDSS EDR BALQSO Catalog --- Objects with $1.7 \le z \le 4.2$ \label{tab:tab1}}
\tablehead{
\colhead{} &
\colhead{} &
\colhead{} &
\colhead{} &
\multicolumn{3}{c}{Target\tablenotemark{a}} &
\colhead{} &
\colhead{} &
\colhead{} &
\colhead{} &
\colhead{} &
\colhead{} &
\colhead{} &
\multicolumn{3}{c}{Classification\tablenotemark{d}} &
\colhead{} \\
\cline{5-7} 
\cline{15-17} 
\multicolumn{7}{c}{} &
\colhead{$f_{\rm 20 cm}$} &
\multicolumn{2}{c}{} &
\colhead{Dereddened} &
\colhead{C IV} &
\colhead{C IV} &
\colhead{Mg II} &
\multicolumn{3}{c}{} &
\colhead{Sub-} \\
\colhead{SDSS Object} &
\colhead{Plate} &
\colhead{Fiber} &
\colhead{MJD} &
\colhead{F} &
\colhead{E} &
\colhead{S} &
\colhead{(mJy)} &
\colhead{Redshift} &
\colhead{$i^*$} &
\colhead{$M_{i^*}$} &
\colhead{$BI$\tablenotemark{b}} &
\colhead{$BI$\tablenotemark{c}} &
\colhead{$BI$\tablenotemark{b}} &
\colhead{1} & 
\colhead{2} &
\colhead{3} &
\colhead{sample} \\
\colhead{(1)} &
\colhead{(2)} &
\colhead{(3)} &
\colhead{(4)} &
\colhead{(5)} &
\colhead{(6)} &
\colhead{(7)} &
\colhead{(8)} &
\colhead{(9)} &
\colhead{(10)} &
\colhead{(11)} &
\colhead{(12)} &
\colhead{(13)} &
\colhead{(14)} &
\colhead{(15)} &
\colhead{(16)} &
\colhead{(17)} &
\colhead{(18)}  
}
\startdata
000056.89$-$010409.8 & 387 &  98 & 51791 & 1 & 1 & 0 &  0.00 & 2.111 & 19.09 & $-$26.34 &  1560 &   300 &    \nodata &  H &  H &      Hi &        Hi \\
001025.90+005447.6 & 389 & 332 & 51795 & 1 & 1 & 1 &  0.00 & 2.845   & 18.78 & $-$27.26 &  4405 &  2163 &    \nodata &  H &  H &     Hiz &        Hi \\
001438.28$-$010750.2 & 389 & 211 & 51795 & 1 & 1 & 1 &  1.41 & 1.813 & 18.33 & $-$26.78 &     0 &    13 &    \nodata &  N &  H &      Hi &        Hi \\
001502.26+001212.4 & 389 & 465 & 51795 & 1 & 1 & 1 &  0.00 & 2.857   & 18.87 & $-$27.17 &   938 &   809 &    \nodata &  H &  H &     Hiz &        Hi \\
001824.96+001525.8 & 390 & 394 & 51900 & 0 & 1 & 0 &  0.00 & 2.430   & 19.55 & $-$26.16 &  3347 &  3468 &    \nodata &  H &  H &     Loz &        Lo \\
002127.88+010420.2 & 390 & 443 & 51900 & 1 & 1 & 1 &  0.00 & 1.829   & 18.01 & $-$27.12 &    40 &    17 &    \nodata &  H &  H &      Hi &        Hi \\
003551.98+005726.3 & 392 & 449 & 51793 & 1 & 1 & 0 &  0.00 & 1.905   & 18.76 & $-$26.46 &  1731 &  1148 &    \nodata &  H &  H &      Hi &        Hi \\
004041.39$-$005537.3 & 393 & 298 & 51794 & 1 & 1 & 0 &  0.00 & 2.092 & 17.91 & $-$27.50 &     0 &    14 &    \nodata &  N &  H &      no &        Hi \\
004118.58+001742.5 & 392 & 631 & 51793 & 1 & 1 & 1 &  0.00 & 1.764   & 18.41 & $-$26.64 &  1384 &    \nodata &    \nodata &  H &  \nodata &      Hi &        Hi \\
004323.43$-$001552.5 & 393 & 181 & 51794 & 0 & 1 & 0 &  0.00 & 2.806 & 17.96 & $-$28.05 &   588 &   340 &    \nodata &  H &  H &     Hiz &        Hi \\
004613.53+010425.7 & 393 & 572 & 51794 & 1 & 1 & 0 &  3.04 & 2.152   & 17.73 & $-$27.73 &  3151 &  2608 &    \nodata &  H &  H &      Hi &        Hi \\
004732.72+002111.4 & 393 & 588 & 51794 & 1 & 1 & 0 &  0.00 & 2.879   & 18.60 & $-$27.47 &   369 &   235 &    \nodata &  H &  H &     Hiz &        Hi \\
005001.81+002620.0 & 394 & 425 & 51876 & 1 & 1 & 0 &  0.00 & 1.936   & 18.93 & $-$26.32 &     0 &    26 &    \nodata &  N &  H &      no &        Hi \\
005355.15$-$000309.3 & 395 & 352 & 51783 & 1 & 1 & 1 &  0.00 & 1.715 & 17.93 & $-$27.07 &  1088 &    \nodata &    \nodata &  H &  \nodata &      Hi &        Hi \\
005419.99+002727.9 & 394 & 514 & 51876 & 0 & 1 & 0 &  0.00 & 2.522   & 18.13 & $-$27.66 &   786 &   802 &    \nodata &  H &  H &     Hiz &        Hi \\
005703.39+001408.1 & 395 & 394 & 51783 & 1 & 1 & 0 &  0.00 & 3.036   & 19.43 & $-$26.74 &  3124 &  2504 &    \nodata &  H &  H &     Hiz &        Hi \\
005830.16+005130.0 & 395 & 446 & 51783 & 1 & 1 & 0 &  0.00 & 1.833   & 18.97 & $-$26.16 &    19 &    73 &    \nodata &  H &  H &     Hi? &        Hi \\
005837.30$-$003553.7 & 395 & 148 & 51783 & 1 & 1 & 1 &  0.00 & 1.810 & 18.97 & $-$26.14 &  4075 &  3166 &    \nodata &  H &  H &      Hi &        Hi \\
010241.04$-$004208.8 & 396 & 261 & 51816 & 1 & 1 & 1 &  0.00 & 1.742 & 18.74 & $-$26.29 &     0\tablenotemark{e} &    \nodata &    \nodata &  N &  \nodata &      Hi &        Hi \\
010336.40$-$005508.7 & 396 & 297 & 51816 & 0 & 1 & 0 &  0.00 & 2.442 & 19.68 & $-$26.05 &    99 &     0 &    \nodata &  H &  N &    Hiz &        Hi \\
010612.21+001920.1 & 396 & 553 & 51816 & 1 & 1 & 0 &  0.00 & 3.110   & 18.16 & $-$28.06 &  2453 &  5698 &    \nodata &  H &  H &      Hi &        Hi \\
010616.05+001523.9 & 396 & 552 & 51816 & 0 & 1 & 0 &  0.00 & 3.050   & 19.73 & $-$26.45 &  2520 &   676 &    \nodata &  H &  H &    Hiz &        Hi \\
011227.60$-$011221.7 & 397 & 122 & 51794 & 1 & 1 & 0 &  0.00 & 1.755 & 17.57 & $-$27.47 &  3033 &    \nodata &    \nodata &  H &  \nodata &      Hi &        Hi \\
011237.35+001929.7 & 397 & 513 & 51794 & 1 & 1 & 0 &  0.00 & 2.695   & 19.53 & $-$26.40 &   760 &   538 &    \nodata &  H &  H &    Hiz? &        Hi \\
011251.12+000921.2 & 397 & 510 & 51794 & 0 & 1 & 1 &  0.00 & 2.865   & 19.64 & $-$26.42 &  6388 &  5852 &    \nodata &  H &  H &     Hiz &        Hi \\
011948.51+004356.0 & 398 & 531 & 51789 & 1 & 1 & 1 &  0.00 & 1.772   & 17.99 & $-$27.07 &     1 &    \nodata &    \nodata &  H &  \nodata &    Hi &        Hi \\
012913.70+011428.0 & 399 & 604 & 51817 & 1 & 1 & 0 &  0.00 & 1.782   & 18.59 & $-$26.48 &   345 &    \nodata &    \nodata &  H &  \nodata &      Hi &        Hi \\
013233.89+011607.1 & 400 & 414 & 51820 & 1 & 1 & 1 &  0.00 & 1.786   & 18.66 & $-$26.42 &   514 &    \nodata &    \nodata &  H &  \nodata &      Hi &        Hi \\
013656.31$-$004623.9 & 400 &   3 & 51820 & 1 & 1 & 1 &  0.00 & 1.709 & 17.83 & $-$27.16 &  1304 &    \nodata &    \nodata &  H &  \nodata &      Hi &        Hi \\
014515.58+002931.0 & 401 & 595 & 51788 & 1 & 1 & 0 &  0.00 & 3.006   & 19.80 & $-$26.35 &  8413 &  8885 &    \nodata &  H &  H &     Loz &        Lo \\
014705.43$-$004148.9 & 402 & 271 & 51793 & 0 & 1 & 0 &  0.00 & 2.108 & 19.01 & $-$26.41 &  4926 &  4017 &    \nodata &  H &  H &      Lo &        Lo \\
014812.80$-$005108.8 & 402 & 299 & 51793 & 1 & 1 & 1 &  3.19 & 1.816 & 18.30 & $-$26.82 &  1217 &  2947 &    \nodata &  H &  H &      Hi &        Hi \\
014836.33+000511.5 & 402 & 429 & 51793 & 1 & 1 & 0 &  0.00 & 3.339   & 19.66 & $-$26.70 &  4065 &  2461 &    \nodata &  H &  H &     hiz &        Hi \\
014905.27$-$011404.9\tablenotemark{f} & 402 & 246 & 51793 & 1 & 1 & 0  & 0.00 & 1.960 & 19.51 & $-$25.76  &  \nodata \tablenotemark{e} & \nodata &    \nodata &  \nodata &  N &   Fe?Lo &      FeLo \\
015024.44+004432.9 & 402 & 485 & 51793 & 1 & 1 & 0 &  0.00 & 1.990   & 18.86 & $-$26.44 &   937 &  1798 &    \nodata &  H &  H &      Hi &        Hi \\
015048.82+004126.2 & 402 & 505 & 51793 & 1 & 1 & 0 &  0.00 & 3.703   & 18.26 & $-$28.31 &   105 &    10 &    \nodata &  H &  H &     Hiz &        Hi \\
020006.31$-$003709.7 & 403 &  70 & 51871 & 0 & 1 & 0 &  0.00 & 2.136 & 18.19 & $-$27.26 &  9550 &  8435 &    \nodata &  H &  H &      Lo &        Lo \\
020529.19$-$000912.9 & 404 & 165 & 51812 & 0 & 1 & 0 &  0.00 & 2.350 & 19.94 & $-$25.71 &     3 &  4535 &    \nodata &  H &  H &     Loz &        Lo \\
021327.25$-$001446.9 & 405 & 197 & 51816 & 0 & 1 & 0 &  0.00 & 2.399 & 19.96 & $-$25.73 &   700 &   116 &    \nodata &  H &  H &    Hiz &        Hi \\
021606.40+011509.5 & 405 & 570 & 51816 & 1 & 1 & 0 &  0.00 & 2.223   & 18.47 & $-$27.06 &  2495 &  1740 &    \nodata &  H &  H &      Hi &        Hi \\
022505.06+001733.2 & 406 & 507 & 51817 & 0 & 1 & 0 &  0.00 & 2.422   & 19.28 & $-$26.43 &  1099 &  1016 &    \nodata &  H &  H &     Loz &        Lo \\
022716.73$-$001317.1 & 406 &  71 & 51817 & 0 & 0 & 1 &  0.00 & 1.980 & 19.73 & $-$25.56 &   561 &    \nodata &    \nodata  & H  & \nodata  & Hi &     Hi \\
022844.09+000217.0 & 406 &  35 & 51817 & 1 & 1 & 0 &  0.00 & 2.716   & 17.68 & $-$28.26 &  1823 &  1539 &    \nodata &  H &  H &  HizLo? &        Lo \\
023139.52+001758.4 & 407 & 483 & 51820 & 1 & 1 & 1 &  0.00 & 2.360   & 18.81 & $-$26.85 &   125 &   261 &    \nodata &  H &  H &     Hiz &        Hi \\
023153.78$-$003232.1 & 407 & 163 & 51820 & 1 & 1 & 0 &  0.00 & 1.721 & 18.63 & $-$26.37 &   108 &    \nodata &    \nodata &  H &  \nodata &     Hi? &        Hi \\
023252.80$-$001351.2 & 407 & 158 & 51820 & 1 & 1 & 1 &  0.00 & 2.025 & 18.62 & $-$26.72 &  2092 &  1076 &    \nodata &  H &  H &      Hi &        Hi \\
023600.49+011113.3 & 408 & 324 & 51821 & 0 & 0 & 1 &  0.00 & 1.826   & 19.73 & $-$25.40 &    28 &     0 &    \nodata &  H &  N &      Hi &        Hi \\
023908.98$-$002121.4 & 408 & 179 & 51821 & 1 & 1 & 0 &  0.00 & 3.777 & 19.49 & $-$27.12 &  2655 &  2758 &    \nodata &  H &  H &     Hiz &        Hi \\
024221.86+004912.7 & 408 & 576 & 51821 & 1 & 1 & 1 &  0.00 & 2.071   & 18.03 & $-$27.36 &   229 &   237 &    \nodata &  H &  H &      Hi &        Hi \\
024304.68+000005.4 & 408 &  80 & 51821 & 1 & 1 & 1 &  0.00 & 2.003   & 17.98 & $-$27.34 &   301 &   397 &    \nodata &  H &  H &      Hi &        Hi \\
024457.18$-$010809.9 & 409 & 282 & 51871 & 1 & 1 & 0 &  0.00 & 3.960 & 18.31 & $-$28.40 &  3630 &    \nodata &    \nodata &  H &  \nodata &     HiZ &        Hi \\
025042.45+003536.7 & 410 & 352 & 51816 & 0 & 1 & 0 &  0.00 & 2.380   & 18.31 & $-$27.37 &  3544 &  3873 &    \nodata &  H &  H &     Loz &        Lo \\
025204.28+003137.0 & 410 & 383 & 51816 & 1 & 1 & 1 &  0.00 & 4.100   & 19.46 & $-$27.32 &   259 &    \nodata &    \nodata &  H &  \nodata &     HiZ &        Hi \\
025247.34+005135.1 & 409 & 612 & 51871 & 1 & 1 & 0 &  0.00 & 2.112   & 18.74 & $-$26.68 &   441 &   489 &    \nodata &  H &  H &      no &        Hi \\
025331.93+001624.7 & 410 & 391 & 51816 & 1 & 1 & 0 &  0.00 & 1.825   & 18.55 & $-$26.58 &   186 &     4 &    \nodata &  H &  H &      Hi &        Hi \\
025516.88+011134.4 & 410 & 450 & 51816 & 0\tablenotemark{h} & 0 & 0  &  0.00 & 2.833 & 18.81 & $-$27.22 &   382 &    42 &    \nodata &  H &  H &      no &        Hi \\
025747.75$-$000503.0 & 410 & 117 & 51816 & 0 & 0 & 1 &  0.00 & 2.192 & 19.79 & $-$25.72 &  2239 &  2156 &    \nodata &  H &  H &      Hi &        Hi \\
030437.22+004729.1 & 411 & 504 & 51817 & 0 & 1 & 0 &  0.00 & 2.425   & 18.88 & $-$26.83 &  1648 &  1380 &    \nodata &  H &  H &     Hiz &        Hi \\
031227.13$-$003446.2 & 412 & 108 & 51931 & 1 & 1 & 1 &  0.00 & 1.772 & 18.85 & $-$26.21 &     0\tablenotemark{e} &    \nodata &    \nodata &  N &  \nodata &      Hi &        Hi \\
031609.83+004043.0 & 413 & 387 & 51929 & 0\tablenotemark{h} & 0 & 0  &  0.00 & 2.902 & 18.33 & $-$27.75 &   140 &   214 &    \nodata &  H &  H &     Hiz &        Hi \\
031828.90$-$001523.2 & 413 & 170 & 51929 & 1 & 1 & 0 &  0.00 & 1.990 & 17.85 & $-$27.45 &   231 &   154 &    \nodata &  H &  H &      Hi &        Hi \\
032118.21$-$010539.9 & 413 &  48 & 51929 & 0 & 1 & 0 & \nodata & 2.412 & 17.45 & $-$28.25 &   \nodata&   537 &    \nodata &  \nodata &  H &  Fe?Loz &      FeLo \\
032125.79+001359.3 & 413 & 551 & 51929 & 1 & 1 & 1 & \nodata & 1.704   & 18.89 & $-$26.09 &  1587 &    \nodata &    \nodata &  H &  \nodata &     Hi? &        Hi \\
032701.43$-$002207.2 & 414 & 152 & 51869 & 0 & 1 & 0 & \nodata & 2.325 & 18.58 & $-$27.05 &   710 &   573 &    \nodata &  H &  H &     Hiz &        Hi \\
033048.51$-$002819.6 & 415 & 268 & 51810 & 0 & 1 & 0 & \nodata & 1.779 & 18.79 & $-$26.28 &  5548 &    \nodata &    \nodata &  H &  \nodata &      Hi &        Hi \\
033335.03$-$004927.0 & 415 & 138 & 51810 & 1 & 1 & 1 & \nodata & 1.776 & 18.61 & $-$26.46 &   299 &    \nodata &    \nodata &  H &  \nodata &      no &        Hi \\
094745.27$-$004113.2 & 266 &   5 & 51630 & 1 & 1 & 0 &  4.67 & 2.835 & 18.85 & $-$27.18 &   690 &   419 &    \nodata &  H &  H &     Hiz &        Hi \\
095048.48$-$000017.7 & 267 & 224 & 51608 & 0 & 1 & 0 &  0.00 & 1.876 & 19.46 & $-$25.72 &   194 &     0 &   660 &  H &  N &      Lo &        Lo \\
100809.63$-$000209.9 & 269 &  31 & 51910 & 0 & 1 & 0 &  0.00 & 2.561 & 19.35 & $-$26.48 &    56 &    19 &    \nodata &  H &  H &      no &        Hi \\
102258.15$-$004052.1 & 272 & 219 & 51941 & 1 & 1 & 0 &  0.00 & 1.758 & 18.26 & $-$26.79 &   998 &    \nodata &    \nodata &  H &  \nodata &      no &        Hi \\
102517.58+003422.0 & 272 & 501 & 51941 & 1 & 1 & 0 &  0.00 & 1.888   & 18.06 & $-$27.13 &   993 &   856 &    \nodata &  H &  H &      Hi &        Hi \\
102527.43$-$000519.1 & 272 & 156 & 51941 & 0 & 1 & 1 &  0.00 & 1.897 & 18.52 & $-$26.69 &   149 &   338 &    \nodata &  H &  H &    Hi &        Hi \\
103544.96$-$002924.8 & 273 &  69 & 51957 & 1 & 1 & 0 &  0.00 & 2.070 & 18.64 & $-$26.75 &     0 &   567 &    \nodata &  N &  H &    Hi &        Hi \\
104109.85+001051.8 & 274 & 482 & 51913 & 1 & 1 & 0 &  0.00 & 2.250   & 19.03 & $-$26.53 &  1913 &   974 &    \nodata &  H &  H &      Hi &        Hi \\
104130.17+000118.8 & 274 & 179 & 51913 & 1 & 1 & 0 &  0.00 & 2.068   & 17.87 & $-$27.51 &   672 &   705 &     0 &  H &  H &   HiLo? &        Lo \\
104152.61$-$001102.1 & 274 & 159 & 51913 & 1 & 1 & 1 &  0.00 & 1.703 & 18.83 & $-$26.14 &  1588 &    \nodata &    \nodata &  H &  \nodata &      Hi &        Hi \\
104233.86+010206.3 & 275 & 321 & 51910 & 1 & 1 & 1 &  0.00 & 2.123   & 18.39 & $-$27.05 &   401 &     0 &    \nodata &  H &  N &      Hi &        Hi \\
104841.02+000042.8 & 276 & 310 & 51909 & 1 & 1 & 0 &  0.00 & 2.022   & 18.75 & $-$26.59 &  1176 &   840 &    \nodata &  H &  H &      Hi &        Hi \\
105621.65+001243.5 & 276 & 551 & 51909 & 0 & 1 & 0 &  0.00 & 1.709   & 19.89 & $-$25.10 &   333 &    \nodata &    \nodata &  H &  \nodata &     Hi? &        Hi \\
110041.19+003631.9 & 277 & 437 & 51908 & 1 & 1 & 1 &  0.00 & 2.017   & 18.18 & $-$27.15 &  4687 &  3985 &    \nodata &  H &  H &      Hi &        Hi \\
110533.17$-$005848.1 & 278 & 286 & 51900 & 1 & 1 & 1 &  0.00 & 1.742 & 18.34 & $-$26.68 &   452 &    \nodata &    \nodata &  H &  \nodata &     Hi? &        Hi \\
110623.52$-$004326.0 & 278 & 251 & 51900 & 0 & 1 & 0 &  0.00 & 2.450 & 18.08 & $-$27.65 &  4034 &  3828 &    \nodata &  H &  H &  HizLo? &        Lo \\
110728.45$-$005122.6 & 278 & 214 & 51900 & 0 & 1 & 0 &  0.00 & 1.730 & 18.94 & $-$26.07 &  2815 &    \nodata &   663 &  H &  \nodata &      Lo &        Lo \\
110736.67+000329.4 & 278 & 271 & 51900 & 1 & 1 & 1 &  0.00 & 1.740   & 17.93 & $-$27.09 &   123 &    \nodata &    \nodata &  H &  \nodata &      Hi &        Hi \\
110838.76$-$005533.7 & 278 & 123 & 51900 & 1 & 1 & 1 &  0.00 & 1.796 & 17.71 & $-$27.38 &     0\tablenotemark{e} &    \nodata &    \nodata &  N &  \nodata &      Hi &        Hi \\
112602.80+003418.3 & 281 & 432 & 51614 & 1 & 1 & 1 &  0.00 & 1.783   & 17.94 & $-$27.14 &  1232 &    \nodata &    \nodata &  H &  \nodata &      Hi &        Hi \\
113212.92+010441.3 & 282 & 330 & 51658 & 1 & 1 & 1 &  0.00 & 2.250   & 19.07 & $-$26.49 &   \nodata \tablenotemark{e} & \nodata &    \nodata &  \nodata &  N &  Fe?Loz &      FeLo \\
113544.33+001118.6 & 282 & 510 & 51658 & 0 & 1 & 0 &  0.00 & 1.723   & 19.40 & $-$25.61 &  3379 &    \nodata &    \nodata &  H &  \nodata &      Hi &        Hi \\
113621.05+005021.2 & 282 & 535 & 51658 & 1 & 1 & 0 &  0.00 & 3.439   & 17.85 & $-$28.57 &   326 &     0 &    \nodata &  H &  N &     Hiz &        Hi \\
114056.80$-$002329.9 & 283 & 315 & 51959 & 0 & 1 & 0 &  0.00 & 3.605 & 19.53 & $-$26.99 &  1637 &  1861 &    \nodata &  H &  H &  HizLo? &        Lo \\
114534.51$-$004338.6 & 283 & 165 & 51959 & 1 & 1 & 1 &  7.52 & 1.748 & 18.92 & $-$26.11 &   467 &    \nodata &    \nodata &  H &  \nodata &     no? &        Hi \\
115031.02$-$004403.1 & 284 & 261 & 51943 & 0 & 1 & 1 &  0.00 & 2.391 & 20.03 & $-$25.65 &  1494 &  1390 &    \nodata &  H &  H &     Hiz &        Hi \\
120657.01$-$002537.8 & 286 & 160 & 51999 & 1 & 1 & 1 &  0.00 & 2.005 & 19.03 & $-$26.29 &   110 &   158 &    \nodata &  H &  H &      Hi &        Hi \\
120834.84+002047.7 & 286 & 598 & 51999 & 0 & 1 & 0 &  0.00 & 2.708   & 18.21 & $-$27.73 &  1646 &   195 &    \nodata &  H &  H &     Hiz &        Hi \\
120957.19$-$002302.0 & 287 & 268 & 52023 & 1 & 1 & 1 &  0.00 & 1.861 & 18.59 & $-$26.58 &   465 &   634 &    \nodata &  H &  H &      Hi &        Hi \\
121323.94+010414.7 & 287 & 527 & 52023 & 0 & 0 & 1 & 21.54 & 2.836   & 20.10 & $-$25.93 &  1968 &   939 &    \nodata &  H &  H &     Hiz &        Hi \\
121549.81$-$003432.2 & 288 & 266 & 52000 & 1 & 1 & 0 &  0.00 & 2.705 & 17.13 & $-$28.81 &  3598 &  4340 &    \nodata &  H &  H &     Hiz &        Hi \\
121633.90+010732.8 & 287 & 604 & 52023 & 1 & 1 & 0 &  0.00 & 2.018   & 18.44 & $-$26.89 &  2200 &  1191 &    \nodata &  H &  H &      Hi &        Hi \\
121803.28+001236.8 & 287 & 624 & 52023 & 1 & 1 & 1 &  0.00 & 2.010   & 18.57 & $-$26.75 &   269 &    30 &    \nodata &  H &  H &      no &        Hi \\
122228.39$-$011011.0 & 288  &  2 & 52000 & 1 & 1 & 0 &  0.00 & 2.284 & 19.03 & $-$26.55 &  678  &    0  &   \nodata &   H &  N &  HizLo? & Lo \\
122848.21$-$010414.5 & 289 &  88 & 51990 & 1 & 1 & 1 & 29.36 & 2.654 & 17.80 & $-$28.10 &   397 &     3 &    \nodata &  H &  H &   Hiz &        Hi \\
122944.93+004253.0 & 289 & 573 & 51990 & 1 & 1 & 0 &  0.00 & 2.883   & 19.56 & $-$26.51 &  1168 &  2168 &    \nodata &  H &  H &     Hiz &        Hi \\
123056.58$-$005306.3 & 289 &  17 & 51990 & 1 & 1 & 0 &  0.00 & 2.162 & 18.72 & $-$26.76 &   652 &  1301 &    \nodata &  H &  H &      Hi &        Hi \\
123124.71+004719.1 & 289 & 610 & 51990 & 1 & 1 & 1 &  0.00 & 1.720   & 19.03 & $-$25.97 &  3134 &    \nodata &    \nodata &  H &  \nodata &      Hi &        Hi \\
123525.27$-$003653.8 & 290 & 186 & 51941 & 1 & 1 & 1 &  0.00 & 1.950 & 19.03 & $-$26.23 &  1076 &   631 &    \nodata &  H &  H &      Hi &        Hi \\
123723.86$-$004210.0 & 290 & 145 & 51941 & 0 & 0 & 1 &  1.04 & 1.819 & 19.27 & $-$25.85 &  2220 &   107 &    \nodata &  H &  H &      Hi &        Hi \\
123824.90+001834.5 & 290 & 510 & 51941 & 1 & 1 & 1 &  0.00 & 2.154   & 19.06 & $-$26.40 &   220 &     0 &    \nodata &  H &  N &     Hi &        Hi \\
123947.61+002516.2 & 291 & 342 & 51928 & 0 & 1 & 0 &  0.00 & 1.869   & 19.44 & $-$25.73 &  7299 &  5875 &    \nodata &  H &  H &      Hi &        Hi \\
124551.45+010504.9 & 291 & 612 & 51928 & 1 & 1 & 1 &  0.00 & 2.808   & 17.88 & $-$28.13 &  3512 &  3145 &    \nodata &  H &  H &     Hiz &        Hi \\
124720.27$-$011343.1 & 292 & 292 & 51609 & 1 & 1 & 1 &  0.00 & 2.283 & 18.93 & $-$26.66 &   336 &     0 &    \nodata &  H &  N &     Hiz &        Hi \\
125241.55$-$002040.6 & 292 & 117 & 51609 & 0 & 1 & 1 &  0.00 & 2.898 & 18.47 & $-$27.61 &  2524 &  1692 &    \nodata &  H &  H &  HizLo? &        Lo \\
130035.29$-$003928.4 & 293 & 104 & 51689 & 0 & 1 & 0 &  0.00 & 3.630 & 19.15 & $-$27.38 &   853 &   778 &    \nodata &  H &  H &    Hiz &        Hi \\
130058.13+010551.5 & 293 & 563 & 51689 & 1 & 1 & 1 &  0.00 & 1.903   & 18.97 & $-$26.24 &  3848 &  3559 &    \nodata &  H &  H &      Hi &        Hi \\
130136.13+000157.8 & 293 &  79 & 51689 & 1 & 1 & 0 &  0.00 & 1.784   & 17.41 & $-$27.67 &  5898 &    \nodata &    \nodata &  H &  \nodata &      Hi &        Hi \\
130147.88$-$003817.3 & 293 &  65 & 51689 & 0 & 1 & 0 &  0.00 & 2.710 & 18.86 & $-$27.08 &  2600 &     0 &    \nodata &  H &  N &   Hiz &        Hi \\
130221.80$-$004638.1 & 293 &  12 & 51689 & 0 & 1 & 0 &  0.00 & 2.701 & 18.64 & $-$27.30 &  1773 &   547 &    \nodata &  H &  H &     Hiz &        Hi \\
130348.94+002010.4 & 294 & 390 & 51986 & 1 & 1 & 0 &  1.08 & 3.655   & 18.68 & $-$27.87 &  1425 &   831 &    \nodata &  H &  H &  HizLo? &        Lo \\
130424.00$-$003757.2 & 294 & 264 & 51986 & 1 & 1 & 0 &  0.00 & 3.035 & 18.23 & $-$27.95 &  1065 &   757 &    \nodata &  H &  H &     Hiz &        Hi \\
130506.70+001908.5 & 294 & 393 & 51986 & 1 & 1 & 0 &  0.00 & 1.913   & 18.85 & $-$26.37 &   509 &   673 &    \nodata &  H &  H &      Hi &        Hi \\
131010.75$-$003007.2 & 294 &  72 & 51986 & 1 & 0 & 1 &  2.59 & 2.630 & 18.87 & $-$27.01 &  4841 & \nodata & \nodata &  H &  N &     Loz &        Lo \\
131333.01$-$005114.3 & 295 & 170 & 51985 & 1 & 1 & 0 &  0.00 & 2.949 & 19.11 & $-$27.00 &  3794 &  2182 &    \nodata &  H &  H &     Loz &        Lo \\
131714.21+010013.0 & 296 & 321 & 51665 & 1 & 1 & 0 &  0.00 & 2.691   & 18.06 & $-$27.87 &  3240 &  2540 &    \nodata &  H &  H &     Hiz &        Hi \\
131853.45+002211.4 & 296 & 386 & 51665 & 1 & 1 & 1 &  0.00 & 2.079   & 18.57 & $-$26.82 &  1473 &  2226 &    \nodata &  H &  H &      Hi &        Hi \\
132139.86$-$004151.9\tablenotemark{g} & 296 & 147 & 51665 & 1 & 0 & 1 &  4.07 & 3.080 & 18.65 & $-$27.56 &  2401 &  5158 &    \nodata &  H &  H &   FeLoz &      FeLo \\
132304.58$-$003856.5\tablenotemark{g} & 296 &  67 & 51665 & 1 & 1 & 0 &  8.94 & 1.828 & 17.82 & $-$27.31 &   287 &   273 &    \nodata &  H &  H &      Hi &        Hi \\
132742.92+003532.6 & 297 & 504 & 51959 & 1 & 1 & 0 &  0.00 & 1.876   & 18.29 & $-$26.89 &   671 &    63 &    \nodata &  H &  H &      Hi &        Hi \\
134145.13$-$003631.0 & 299 & 172 & 51671 & 0 & 1 & 0 &  0.00 & 2.205 & 18.50 & $-$27.02 &   870 &  1051 &    \nodata &  H &  H &    FeLo &      FeLo \\
134544.55+002810.8 & 300 & 426 & 51666 & 1 & 1 & 1 &  0.00 & 2.516   & 18.48 & $-$27.31 &  1510 &  2017 &    \nodata &  H &  H &     Hiz &        Hi \\
134808.79+003723.2 & 300 & 461 & 51666 & 1 & 1 & 0 &  0.00 & 3.620   & 19.28 & $-$27.25 &  1309 &   986 &    \nodata &  H &  H &     Hiz &        Hi \\
135317.80$-$000501.3 & 300 &  33 & 51666 & 0 & 1 & 0 &  0.00 & 2.320 & 19.04 & $-$26.58 &  9821 & 10190 &    \nodata &  H &  H &     Loz &        Lo \\
135559.04$-$002413.6 & 301 & 267 & 51942 & 1 & 1 & 0 &  0.00 & 2.332 & 18.13 & $-$27.50 &  1525 &   917 &    \nodata &  H &  H &     Hiz &        Hi \\
135702.92+003824.4 & 301 & 414 & 51942 & 0 & 1 & 0 &  0.00 & 2.317   & 19.47 & $-$26.15 &   415 &   618 &    \nodata &  H &  H &    Hiz &        Hi \\
135721.77+005501.1 & 301 & 408 & 51942 & 1 & 1 & 1 &  2.75 & 2.001   & 17.98 & $-$27.33 &   235 &   293 &    \nodata &  H &  H &     Hi? &        Hi \\
140918.72+004824.3 & 302 & 535 & 51688 & 1 & 1 & 0 &  0.00 & 2.000   & 18.44 & $-$26.87 &  1411 &   633 &    \nodata &  H &  H &      Hi &        Hi \\
141332.35$-$004909.5 & 303 & 213 & 51615 & 1 & 1 & 0 &  0.00 & 4.140 & 19.19 & $-$27.61 &   444 &    \nodata &    \nodata &  H &  \nodata &    HiZ &        Hi \\
142050.34$-$002553.1 & 304 & 123 & 51609 & 1 & 1 & 0 &  0.00 & 2.103 & 18.62 & $-$26.80 &  3442 &  2016 &    \nodata &  H &  H &   HiLo? &        Lo \\
142232.38$-$003043.9 & 304 &  81 & 51609 & 0 & 1 & 0 &  0.00 & 2.711 & 19.28 & $-$26.66 &   873 &   308 &    \nodata &  H &  H &     Hiz &        Hi \\
142647.80+002739.9 & 305 & 476 & 51613 & 1 & 1 & 0 &  0.00 & 3.711   & 19.28 & $-$27.30 &   197 &   126 &    \nodata &  H &  H &     Hiz &        Hi \\
143022.47$-$002045.2 & 306 & 221 & 51637 & 0 & 1 & 0 &  0.00 & 2.544 & 19.35 & $-$26.46 &  1957 &  2033 &    \nodata &  H &  H &     Hiz &        Hi \\
143054.03$-$003627.3 & 306 & 225 & 51637 & 1 & 1 & 0 &  0.00 & 3.710 & 19.20 & $-$27.37 &  9064 &  8737 &    \nodata &  H &  H &  HizLo? &        Lo \\
143307.40+003319.0 & 306 & 546 & 51637 & 0 & 1 & 1 &  0.00 & 2.745   & 19.11 & $-$26.85 &  1832 &   517 &    \nodata &  H &  H &     Hiz &        Hi \\
143641.24+001558.9 & 306 & 628 & 51637 & 1 & 1 & 1 &  0.00 & 1.867   & 18.33 & $-$26.84 &  3890 &  2297 &    \nodata &  H &  H &      Hi &        Hi \\
144256.86$-$004501.0 & 307 &  11 & 51663 & 1 & 1 & 0 &  0.00 & 2.226 & 18.17 & $-$27.36 &   815 &    29 &    \nodata &  H &  H &    Hi &        Hi \\
144959.97+003225.3 & 309 & 357 & 51994 & 1 & 1 & 0 &  0.00 & 1.709   & 18.94 & $-$26.05 &  1530 &    \nodata &    \nodata &  H &  \nodata &      Hi &        Hi \\
145045.42$-$004400.3 & 308 &  70 & 51662 & 1 & 1 & 1 &  0.00 & 2.078 & 18.27 & $-$27.12 &   238 &     0 &    \nodata &  H &  N &      Hi &        Hi \\
145411.91+000341.5 & 309 & 472 & 51994 & 1 & 1 & 0 &  0.00 & 1.712   & 18.95 & $-$26.04 &  1677 &    \nodata &    \nodata &  H &  \nodata &    Hi &        Hi \\
145913.72+000215.8 & 310 & 305 & 51990 & 1 & 1 & 1 &  0.00 & 1.910   & 18.55 & $-$26.67 &   356 &   266 &    \nodata &  H &  H &      Hi &        Hi \\
150033.52+003353.7 & 310 & 388 & 51990 & 1 & 1 & 0 &  0.00 & 2.451   & 18.02 & $-$27.71 &  4257 &  3633 &    \nodata &  H &  H &     Hiz &        Hi \\
150114.37$-$005340.9 & 310 & 241 & 51990 & 1 & 1 & 0 &  0.00 & 3.279 & 19.48 & $-$26.85 &   281 &     0 &    \nodata &  H &  N &     Hiz &        Hi \\
150206.66$-$003606.9 & 310 & 236 & 51990 & 1 & 1 & 0 &  9.87 & 2.202 & 18.49 & $-$27.02 &   406 &     0 &    \nodata &  H &  N &      Hi &        Hi \\
151636.79+002940.4\tablenotemark{g} & 312 & 434 & 51689 & 1 & 0 & 0  &  2.19 & 2.240 & 17.25 & $-$28.29 &  4035 & \nodata &    \nodata &  H &  \nodata &      Lo &        Lo \\
152348.99$-$004701.8 & 313 & 213 & 51673 & 1 & 1 & 0 &  0.00 & 3.293 & 18.21 & $-$28.13 &  2913 &  4333 &    \nodata &  H &  H &     Hiz &        Hi \\
152913.85$-$001013.8 & 313 &  32 & 51673 & 1 & 1 & 1 &  0.00 & 2.073 & 18.23 & $-$27.16 &   209 &     0 &    \nodata &  H &  N &      Hi &        Hi \\
170056.85+602639.8 & 353 & 336 & 51703 & 1 & 1 & 1 & \nodata & 2.125   & 18.51 & $-$26.93 &  1400 &  1461 &    \nodata &  H &  H &      Hi &        Hi \\
170633.05+615715.1 & 351 & 555 & 51780 & 1 & 1 & 1 & \nodata & 2.008   & 18.56 & $-$26.76 &   171 &    28 &    \nodata &  H &  H &      Hi &        Hi \\
170720.18+613025.5 & 351 & 147 & 51780 & 1 & 1 & 1 & \nodata & 1.744   & 18.74 & $-$26.29 &  2655 &    \nodata &    \nodata &  H &  \nodata &      Hi &        Hi \\
170903.06+594530.7 & 353 & 517 & 51703 & 1 & 1 & 1 & \nodata & 1.708   & 18.64 & $-$26.34 &  4936 &    \nodata &    \nodata &  H &  \nodata &      Hi &        Hi \\
170931.00+630357.1 & 352 & 311 & 51694 & 0 & 1 & 0 & \nodata & 2.402   & 17.25 & $-$28.44 &   \nodata&  1232 &    \nodata &  \nodata &  H &  Fe?Loz &      FeLo \\
170951.03+570313.7 & 355 & 310 & 51788 & 0 & 1 & 0 &  2.21 & 2.547   & 19.87 & $-$25.94 &   528 &     0 &    \nodata &  H &  N &   Hiz &        Hi \\
171652.35+590200.2 & 353 &  63 & 51703 & 1 & 1 & 1 & \nodata & 2.369   & 18.65 & $-$27.01 &   620 &   716 &    \nodata &  H &  H &     Hiz &        Hi \\
171831.73+595309.3 & 353 & 624 & 51703 & 1 & 1 & 0 & \nodata & 1.832   & 18.64 & $-$26.49 &  1406 &  1423 &    \nodata &  H &  H &      Hi &        Hi \\
171944.76+554408.3 & 367 & 436 & 51997 & 1 & 1 & 0 &  0.00 & 3.886   & 19.69 & $-$26.98 &   205 &    \nodata &    \nodata &  H &  \nodata &   Hiz &        Hi \\
171949.92+532132.8 & 359 & 308 & 51821 & 1 & 1 & 1 &  0.00 & 1.777   & 18.00 & $-$27.07 &  4903 &    \nodata &    \nodata &  H &  \nodata &      Hi &        Hi \\
172001.31+621245.7 & 352 & 125 & 51694 & 1 & 1 & 1 & \nodata & 1.762   & 18.70 & $-$26.35 &  3290 &    \nodata &    \nodata &  H &  \nodata &      Hi &        Hi \\
172012.40+545601.0 & 367 & 184 & 51997 & 1 & 1 & 0 &  0.00 & 2.099   & 18.24 & $-$27.18 &  1249 &    44 &    \nodata &  H &  H &      Hi &        Hi \\
172310.22+573835.3 & 366 & 250 & 52017 & 0 & 0 & 1 &  0.00 & 1.780   & 20.00 & $-$25.07 &  1817 &    \nodata &    \nodata &  H &  \nodata &      Hi &        Hi \\
172341.09+555340.5\tablenotemark{f} & 367 & 506 & 51997 & 1 & 1 & 1 &  0.00 & 2.113 & 18.47 & $-$26.96 &  3497 &     \nodata &    \nodata &  H &  N &    FeLo &      FeLo \\
172413.28+571046.7 & 358 & 335 & 51818 & 1 & 1 & 0 &  0.00 & 2.815   & 17.88 & $-$28.14 &  5806 &  6826 &    \nodata &  H &  H &     Hiz &        Hi \\
172656.65+535308.4 & 359 & 499 & 51821 & 0 & 1 & 0 &  0.00 & 2.905   & 19.61 & $-$26.47 &  3770 &  3388 &    \nodata &  H &  H &     Hiz &        Hi \\
173049.10+585059.5\tablenotemark{f} & 366 & 558 & 52017 & 0 & 0 & 1 & \nodata & 1.980 & 20.76 & $-$24.54 &     0\tablenotemark{e} & \nodata &    \nodata &  N &  N &    FeLo &      FeLo \\
173218.35+602014.0 & 354 &  69 & 51792 & 0 & 1 & 0 & \nodata & 2.356   & 19.80 & $-$25.86 &   534 &   104 &    \nodata &  H &  H &    Hiz &        Hi \\
173315.24+584814.3 & 366 & 593 & 52017 & 0 & 0 & 1 & \nodata & 1.907   & 19.68 & $-$25.54 &   374 &   179 &    \nodata &  H &  H &      Hi &        Hi \\
173722.98+572116.7 & 358 & 564 & 51818 & 1 & 1 & 0 &  0.00 & 1.944   & 19.00 & $-$26.25 &   293 &   163 &    \nodata &  H &  H &      Hi &        Hi \\
173802.91+535047.2 & 360 & 132 & 51816 & 1 & 1 & 1 &  0.00 & 1.870   & 18.18 & $-$27.00 &  2095 &  1305 &    \nodata &  H &  H &      Hi &        Hi \\
173911.52+565550.9 & 358 & 592 & 51818 & 1 & 1 & 0 &  0.00 & 1.772   & 18.90 & $-$26.16 &   919 &    \nodata &    \nodata &  H &  \nodata &    Hi &        Hi \\
173935.27+575201.7 & 358 & 605 & 51818 & 1 & 1 & 0 &  0.00 & 3.222   & 18.84 & $-$27.45 &   473 &    42 &    \nodata &  H &  H &     Hiz &        Hi \\
232205.46+004550.9 & 383 & 388 & 51818 & 0 & 0 & 1 &  0.00 & 1.820   & 19.67 & $-$25.45 &   222 &    28 &    \nodata &  H &  H &      Hi &        Hi \\
233934.42$-$002932.6 & 385 & 221 & 51877 & 0 & 1 & 0 &  0.00 & 2.010 & 19.70 & $-$25.63 &  1242 &  2376 &    \nodata &  H &  H &      Hi &        Hi \\
234506.32+010135.5 & 385 & 617 & 51877 & 1 & 1 & 1 &  0.00 & 1.794   & 18.64 & $-$26.45 &  2488 &    \nodata &    \nodata &  H &  \nodata &      Hi &        Hi \\
234812.39+002939.5 & 386 & 388 & 51788 & 1 & 1 & 0 &  0.00 & 1.947   & 18.91 & $-$26.35 &  5100 &  5289 &  1002 &  H &  H &   HiLo? &        Lo \\
235224.13$-$000951.0 & 386 & 180 & 51788 & 1 & 1 & 0 &  0.00 & 2.769 & 19.17 & $-$26.81 &  1157 &  1556 &    \nodata &  H &  H &     Hiz &        Hi \\
235408.59$-$001615.1 & 386 & 106 & 51788 & 0 & 1 & 0 &  0.00 & 1.770 & 18.71 & $-$26.35 & 10683 &    \nodata &   715 &  H &  \nodata &      Lo &        Lo \\
235546.14$-$002342.9 & 386 &  28 & 51788 & 1 & 1 & 0 &  0.00 & 3.254 & 19.11 & $-$27.20 &     7 &    12 &    \nodata &  H &  H &      no &        Hi \\
235628.96$-$003602.0 & 387 & 315 & 51791 & 1 & 1 & 0 &  0.00 & 2.940 & 18.61 & $-$27.50 &   428 &   753 &    \nodata &  H &  H &     Hiz &        Hi \\
235859.46$-$002426.2 & 387 & 181 & 51791 & 1 & 1 & 1 &  0.00 & 1.757 & 17.98 & $-$27.07 &  2199 &    \nodata &    \nodata &  H &  \nodata &      Hi &        Hi 
\enddata 
\tablenotetext{a}{Flags for final quasar target selection (F), EDR quasar target selection (E), and EDR serendipity target selection (S): 1 = selected, 0 = not selected.}
\tablenotetext{b}{Balnicity index computed using a fitted EDR composite spectrum as the continuum.}
\tablenotetext{c}{Balnicity index computed using a power law with Gaussian C IV emission line as the continuum.}
\tablenotetext{d}{Classification 1 is determined by the
composite-fitting algorithm: H = HiBAL, N = nonBALQSO.  Classification
2 is determined by the power law + Gaussian algorithm: H = HiBAL, N =
nonBALQSO.  Classification 3 lists the visual classifications as
described in the text.}
\tablenotetext{e}{The BIs of these objects are formally zero (or
unmeasured), but visual inspection suggests that the objects belong in
this table.  For SDSS0149$-$0114 and SDSS1132+0104 a small change in
redshift would yield a positive BI.  In the case of SDSS1730+5850,
\markcite{hal+02}{Hall} {et~al.} (2002) found a BI of $>$10900.}
\tablenotetext{f}{See also \markcite{hal+02}{Hall} {et~al.} (2002).}
\tablenotetext{g}{See also \markcite{mvi+01}{Menou} {et~al.} (2001).}
\tablenotetext{h}{This quasar was selected for spectroscopic follow-up as a star.}
\end{deluxetable}

\clearpage

\begin{deluxetable}{lrrrcccrcccrrrcclc}
\rotate
\tabletypesize{\tiny}
\tablewidth{0pt}
\tablecaption{Supplementary SDSS EDR BALQSO Catalog \label{tab:tab2}}
\tablehead{
\colhead{} &
\colhead{} &
\colhead{} &
\colhead{} &
\multicolumn{3}{c}{Target\tablenotemark{a}} &
\colhead{} &
\colhead{} &
\colhead{} &
\colhead{} &
\colhead{} &
\colhead{} &
\colhead{} &
\multicolumn{3}{c}{Classification\tablenotemark{d}} &
\colhead{} \\
\cline{5-7} 
\cline{15-17} 
\multicolumn{7}{c}{} &
\colhead{$f_{\rm 20 cm}$} &
\multicolumn{2}{c}{} &
\colhead{Dereddened} &
\colhead{C IV} &
\colhead{C IV} &
\colhead{Mg II} &
\multicolumn{3}{c}{} &
\colhead{Sub-} \\
\colhead{SDSS Object} &
\colhead{Plate} &
\colhead{Fiber} &
\colhead{MJD} &
\colhead{F} &
\colhead{E} &
\colhead{S} &
\colhead{(mJy)} &
\colhead{Redshift} &
\colhead{$i^*$} &
\colhead{$M_{i^*}$} &
\colhead{$BI$\tablenotemark{b}} &
\colhead{$BI$\tablenotemark{c}} &
\colhead{$BI$\tablenotemark{b}} &
\colhead{1} & 
\colhead{2} &
\colhead{3} &
\colhead{sample} \\
\colhead{(1)} &
\colhead{(2)} &
\colhead{(3)} &
\colhead{(4)} &
\colhead{(5)} &
\colhead{(6)} &
\colhead{(7)} &
\colhead{(8)} &
\colhead{(9)} &
\colhead{(10)} &
\colhead{(11)} &
\colhead{(12)} &
\colhead{(13)} &
\colhead{(14)} &
\colhead{(15)} &
\colhead{(16)} &
\colhead{(17)} & 
\colhead{(18)} 
}
\startdata
002342.98+010242.8 & 390 & 610 & 51900 & 1 & 1 & 1 &  0.00 & 1.637   & 18.24 & $-$26.66 &    \nodata &    \nodata &    \nodata &  \nodata &  \nodata &      Hi &        Hi \\
010739.80$-$011042.5 & 396 &  41 & 51816 & 1 & 1 & 1 &  0.00 & 1.597 & 18.78 & $-$26.06 &    \nodata &    \nodata &    \nodata &  \nodata &  \nodata &      Hi &        Hi \\
010855.02$-$005747.2 & 397 & 296 & 51794 & 1 & 1 & 0 &  0.00 & 1.673 & 17.37 & $-$27.58 &    \nodata &    \nodata &    \nodata &  \nodata &  \nodata &      Hi &        Hi \\
011913.21+005115.8 & 398 & 453 & 51789 & 1 & 1 & 0 &  0.00 & 1.668   & 18.76 & $-$26.18 &    \nodata &    \nodata &    \nodata &  \nodata &  \nodata &      Hi &        Hi \\
012331.05+011314.6 & 399 & 367 & 51817 & 1 & 1 & 0 &  0.00 & 1.555   & 18.88 & $-$25.91 &    \nodata &    \nodata &    \nodata &  \nodata &  \nodata &      Hi &        Hi \\
012841.87$-$003317.2 & 399 &  26 & 51817 & 1 & 1 & 0 &  3.03 & 1.660 & 18.16 & $-$26.77 &    \nodata &    \nodata &    \nodata &  \nodata &  \nodata &      Hi &        Hi \\
013245.30$-$004610.0 & 400 & 252 & 51820 & 1 & 1 & 0 &  0.00 & 1.475 & 18.32 & $-$26.36 &    \nodata &    \nodata &   194 &  \nodata &  \nodata &      Lo &        Lo \\
014055.58+003908.4 & 401 & 437 & 51788 & 1 & 1 & 1 &  0.00 & 1.492   & 18.79 & $-$25.91 &    \nodata &    \nodata &    \nodata &  \nodata &  \nodata &      Hi &        Hi \\
020105.14+000617.9 & 403 & 593 & 51871 & 1 & 1 & 1 &  0.00 & 1.214   & 17.84 & $-$26.43 &    \nodata &    \nodata &     0 &  \nodata &  \nodata &      Lo &        Lo \\
025204.18+010710.3 & 410 & 321 & 51816 & 0 & 0 & 1 &  0.00 & 1.221   & 19.70 & $-$24.58 &    \nodata &    \nodata &  4093 &  \nodata &  \nodata &      Lo &        Lo \\
030000.56+004828.0\tablenotemark{f} & 410 & 621 & 51816 & 1 & 1 & 0 &  0.00 & 0.892 & 16.38 & $-$27.23 &  \nodata &    \nodata &    \nodata &  \nodata &  \nodata &    FeLo &      FeLo \\
032246.82$-$005148.9 & 414 & 287 & 51869 & 1 & 1 & 1 & \nodata & 1.680 & 18.87 & $-$26.08 &    \nodata &    \nodata &    \nodata &  \nodata &  \nodata &      Hi &        Hi \\
033818.29$-$003710.7 & 416 & 296 & 51811 & 0 & 0 & 1 & \nodata & 1.582 & 19.42 & $-$25.40 &    \nodata &    \nodata &    \nodata &  \nodata &  \nodata &      Hi &        Hi \\
094302.93$-$001310.6 & 266 & 238 & 51630 & 1 & 1 & 1 &  0.00 & 1.590 & 18.38 & $-$26.46 &    \nodata &    \nodata &    \nodata &  \nodata &  \nodata &      Hi &        Hi \\
100021.83$-$010031.9 & 269 & 297 & 51910 & 1 & 1 & 0 &  0.00 & 1.656 & 18.95 & $-$25.97 &    \nodata &    \nodata &    \nodata &  \nodata &  \nodata &      Hi &        Hi \\
105352.86$-$005852.7 & 276 & 139 & 51909 & 1 & 1 & 1 & 24.01 & 1.572 & 17.50 & $-$27.31 &    \nodata &    \nodata &     0 &  \nodata &  \nodata &      Lo &        Lo \\
110826.31+003706.7 & 278 & 435 & 51900 & 1 & 1 & 0 &  0.00 & 4.410   & 19.74 & $-$27.19 &    \nodata &    \nodata &    \nodata &  \nodata &  \nodata &     HiZ &        Hi \\
111249.65+005310.1 & 278 & 619 & 51900 & 1 & 1 & 0 &  0.00 & 1.682   & 17.58 & $-$27.38 &    \nodata &    \nodata &    \nodata &  \nodata &  \nodata &      Hi &        Hi \\
113537.56+004130.1 & 282 & 540 & 51658 & 1 & 1 & 0 &  0.00 & 1.551   & 18.37 & $-$26.41 &    \nodata &    \nodata &    \nodata &  \nodata &  \nodata &      Hi &        Hi \\
114954.93+001255.2 & 284 & 353 & 51943 & 1 & 1 & 0 &  0.00 & 1.597   & 18.08 & $-$26.77 &    \nodata &    \nodata &    \nodata &  \nodata &  \nodata &      Hi &        Hi \\
115357.27$-$002754.0 & 285 & 302 & 51930 & 1 & 1 & 1 &  0.00 & 1.674 & 19.05 & $-$25.90 &    \nodata &    \nodata &    \nodata &  \nodata &  \nodata &      Hi &        Hi \\
115404.13+001419.5\tablenotemark{g} & 284 & 516 & 51943 & 1 & 0 & 0 &  1.46 & 1.610 & 17.77 & $-$27.10 &    \nodata &    \nodata &     0 &  \nodata &  \nodata &      Lo &        Lo \\
120627.62+002335.3 & 286 & 499 & 51999 & 0 & 1 & 0 &  0.00 & 1.111   & 18.64 & $-$25.44 &    \nodata &    \nodata &    \nodata &  \nodata &  \nodata &    FeLo &      FeLo \\
120725.54+010154.8 & 286 & 570 & 51999 & 0 & 1 & 0 &  0.00 & 1.690   & 19.86 & $-$25.10 &    \nodata &    \nodata &    \nodata &  \nodata &  \nodata &      Hi &        Hi \\
121441.42$-$000137.9\tablenotemark{f} & 287 & 514 & 52023 & 1 & 0 & 1 &  1.79 & 1.042 & 18.76 & $-$25.19 &    \nodata &    \nodata &  2590 &  \nodata &  \nodata &      Lo &        Lo \\
121701.50$-$002958.5 & 288 & 264 & 52000 & 1 & 1 & 0 &  0.00 & 1.608 & 19.05 & $-$25.81 &    \nodata &    \nodata &    \nodata &  \nodata &  \nodata &      Hi &        Hi \\
130208.26$-$003731.6\tablenotemark{g} & 293 &  76 & 51689 & 1 & 1 & 0 & 11.24 & 1.672 & 17.60 & $-$27.35 &    \nodata &    \nodata &     0 &  \nodata &  \nodata &      Lo &        Lo \\
145857.57+002621.9 & 310 & 357 & 51990 & 1 & 1 & 1 &  0.00 & 1.554   & 18.79 & $-$26.00 &    \nodata &    \nodata &    \nodata &  \nodata &  \nodata &      Hi &        Hi \\
152839.31$-$002229.3 & 313 &  65 & 51673 & 1 & 1 & 1 &  0.00 & 1.593 & 18.33 & $-$26.51 &    \nodata &    \nodata &    \nodata &  \nodata &  \nodata &      Hi &        Hi \\
171124.23+593121.4 & 353 & 514 & 51703 & 1 & 1 & 0 & \nodata & 1.491   & 18.72 & $-$25.98 &    \nodata &    \nodata &    59 &  \nodata &  \nodata &      Lo &        Lo \\
171330.98+610707.8 & 351 &  40 & 51780 & 1 & 1 & 1 & \nodata & 1.685   & 18.78 & $-$26.18 &    \nodata &    \nodata &    \nodata &  \nodata &  \nodata &      Hi &        Hi \\
171430.12+561523.9 & 367 & 338 & 51997 & 1 & 1 & 0 &  0.00 & 1.678   & 18.51 & $-$26.44 &    \nodata &    \nodata &    \nodata &  \nodata &  \nodata &      Hi &        Hi \\
172308.15+524455.4 & 359 & 207 & 51821 & 0\tablenotemark{e} & 0 & 0 &  2.07 & 1.815 & 17.07 & $-$28.05 &     0 &     0 &    \nodata &  N &  N &     Hi? &        Hi \\
172630.05+615208.3 & 354 & 529 & 51792 & 0 & 0 & 1 & \nodata & 1.591   & 19.83 & $-$25.00 &    \nodata &    \nodata &    \nodata &  \nodata &  \nodata &      Hi &        Hi \\
172718.39+585227.9 & 366 & 507 & 52017 & 1 & 1 & 1 &  0.00 & 1.550   & 18.65 & $-$26.14 &    \nodata &    \nodata &     0 &  \nodata &  \nodata &      Lo &        Lo \\
173221.93+604854.9 & 354 & 638 & 51792 & 1 & 1 & 0 & \nodata & 1.540   & 18.12 & $-$26.65 &    \nodata &    \nodata &    \nodata &  \nodata &  \nodata &      Hi &        Hi \\
173523.03+554611.1 & 360 & 414 & 51816 & 1 & 1 & 0 &  0.00 & 1.588   & 17.42 & $-$27.41 &    \nodata &    \nodata &    \nodata &  \nodata &  \nodata &      Hi &        Hi \\
235238.08+010552.4 & 386 & 524 & 51788 & 0 & 1 & 0 &  0.00 & 2.156   & 17.23 & $-$28.23 &     0 &     0 &    \nodata &  N &  N &      Hi &        Hi \\
235253.51$-$002850.4 & 386 & 167 & 51788 & 1 & 1 & 1 &  0.00 & 1.628 & 17.87 & $-$27.02 &    \nodata &    \nodata &    54 &  \nodata &  \nodata &   HiLo? &        Lo 
\enddata 
\tablenotetext{a}{Flags for final quasar target selection (F), EDR quasar target selection (E), and EDR serendipity target selection (S): 1 = selected, 0 = not selected.}
\tablenotetext{b}{Balnicity index computed using a fitted EDR composite spectrum as the continuum.}
\tablenotetext{c}{Balnicity index computed using a power law with Gaussian C IV emission line as the continuum.}
\tablenotetext{d}{Classification 1 is determined by the composite-fitting algorithm: H = HiBAL, N = nonBALQSO.  Classification 2 is determined by the power law + Gaussian algorithm: H = HiBAL, N = nonBALQSO.  Classification 3 lists the visual classifications as described in the text.}
\tablenotetext{e}{This quasar was selected for spectroscopic follow-up as a galaxy.}
\tablenotetext{f}{See also \markcite{hal+02}{Hall} {et~al.} (2002).}
\tablenotetext{g}{See also \markcite{mvi+01}{Menou} {et~al.} (2001).}
\end{deluxetable}


\begin{thebibliography}{}

\bibitem[{Baldwin} 1977]{bal77}
{Baldwin}, J.~A. 1977, \apj, 214, 679

\bibitem[{Becker}, {White}, {Gregg}, {Brotherton},  {Laurent-Muehleisen}, \& {Arav} 2000]{bwg+00}
{Becker}, R.~H., {White}, R.~L., {Gregg}, M.~D., {Brotherton}, M.~S.,  {Laurent-Muehleisen}, S.~A., \& {Arav}, N. 2000, \apj, 538, 72

\bibitem[{Becker}, {White}, \& {Helfand} 1995]{bwh95}
{Becker}, R.~H., {White}, R.~L., \& {Helfand}, D.~J. 1995, \apj, 450, 559

\bibitem[{Blanton et al.} 2002]{blm+02}
{Blanton et al.} 2002, \aj, submitted

\bibitem[{Boroson} \& {Meyers} 1992]{bm92}
{Boroson}, T.~A. \& {Meyers}, K.~A. 1992, \apj, 397, 442

\bibitem[{Brotherton}, {Tran}, {Becker}, {Gregg},  {Laurent-Muehleisen}, \& {White} 2001]{btb+01}
{Brotherton}, M.~S., {Tran}, H.~D., {Becker}, R.~H., {Gregg}, M.~D.,  {Laurent-Muehleisen}, S.~A., \& {White}, R.~L. 2001, \apj, 546, 775

\bibitem[{Cardelli}, {Clayton}, \& {Mathis} 1989]{ccm89}
{Cardelli}, J.~A., {Clayton}, G.~C., \& {Mathis}, J.~S. 1989, \apj, 345, 245

\bibitem[{Foltz}, {Weymann}, {Peterson}, {Sun},  {Malkan}, \& {Chaffee} 1986]{fwp+86}
{Foltz}, C.~B., {Weymann}, R.~J., {Peterson}, B.~M., {Sun}, L., {Malkan},  M.~A., \& {Chaffee}, F.~H. 1986, \apj, 307, 504

\bibitem[{Fukugita}, {Ichikawa}, {Gunn}, {Doi},  {Shimasaku}, \& {Schneider} 1996]{fig+96}
{Fukugita}, M., {Ichikawa}, T., {Gunn}, J.~E., {Doi}, M., {Shimasaku}, K., \&  {Schneider}, D.~P. 1996, \aj, 111, 1748

\bibitem[{Gunn}, {Carr}, {Rockosi}, {Sekiguchi}, {Berry},  {Elms}, {de Haas}, {Ivezi{\' c}}, {Knapp}, {Lupton}, {Pauls}, {Simcoe},  {Heidtman}, {Schneider}, {Lucinio}, \& {Brinkman} 1998]{gcr+98}
{Gunn}, J.~E., {Carr}, M., {Rockosi}, C., {Sekiguchi}, M., {Berry}, K., {Elms},  B., {de Haas}, E., {Ivezi{\' c}}, {\v Z}., {et al.} 1998, \aj, 116, 3040

\bibitem[{Hall}, {Anderson}, {Strauss}, {York},  {Richards}, {Fan}, {Knapp}, {Schneider}, {Vanden Berk}, {Geballe}, {Bauer},  {Becker}, {Davis}, {Rix}, {Nichol}, {Bahcall}, {Brinkmann}, {Brunner},  {Connolly}, {Csabai}, {Doi}, {Fukugita}, {Gunn}, {Haiman}, {Harvanek},  {Heckman}, {Hennessy}, {Inada}, {Ivezi{\' c}}, {Johnston}, {Kleinman},  {Krolik}, {Krzesinski}, {Kunszt}, {Lamb}, {Long}, {Lupton}, {Miknaitis},  {Munn}, {Narayanan}, {Neilsen}, {Newman}, {Nitta}, {Okamura}, {Pentericci},  {Pier}, {Schlegel}, {Snedden}, {Szalay}, {Thakar}, {Tsvetanov}, {White}, \&  {Zheng} 2002]{hal+02}
{Hall}, P.~B., {Anderson}, S.~F., {Strauss}, M.~A., {York}, D.~G., {Richards},  G.~T., {Fan}, X., {Knapp}, G.~R., {Schneider}, D.~P., {et al.} 2002, \apjs, 141, 267

\bibitem[{Hewett}, {Foltz}, \& {Chaffee} 1995]{hfc+95}
{Hewett}, P.~C., {Foltz}, C.~B., \& {Chaffee}, F.~H. 1995, \aj, 109, 1498

\bibitem[{Hogg}, {Finkbeiner}, {Schlegel}, \&  {Gunn} 2001]{hfs+01}
{Hogg}, D.~W., {Finkbeiner}, D.~P., {Schlegel}, D.~J., \& {Gunn}, J.~E. 2001,  \aj, 122, 2129

\bibitem[{Menou}, {Vanden Berk}, {Ivezi{\' c}}, {Kim},  {Knapp}, {Richards}, {Strateva}, {Fan}, {Gunn}, {Hall}, {Heckman}, {Krolik},  {Lupton}, {Schneider}, {York}, {Anderson}, {Bahcall}, {Brinkmann}, {Brunner},  {Csabai}, {Fukugita}, {Hennessy}, {Kunszt}, {Lamb}, {Munn}, {Nichol}, \&  {Szokoly} 2001]{mvi+01}
{Menou}, K., {Vanden Berk}, D.~E., {Ivezi{\' c}}, {\v Z}., {Kim}, R.~S.~J.,  {Knapp}, G.~R., {Richards}, G.~T., {Strateva}, I., {Fan}, X., {et al.} 2001, \apj, 561, 645

\bibitem[{Mihalas} \& {Binney} 1981]{mb81}
{Mihalas}, D. \& {Binney}, J. 1981, {Galactic Astronomy: Structure and  Kinematics} (San Francisco, CA, W.~H.~Freeman and Co.)

\bibitem[{Nandy}, {Morgan}, {Willis}, {Wilson}, \&  {Gondhalekar} 1981]{nmw+81}
{Nandy}, K., {Morgan}, D.~H., {Willis}, A.~J., {Wilson}, R., \& {Gondhalekar},  P.~M. 1981, \mnras, 196, 955

\bibitem[{Ogle}, {Cohen}, {Miller}, {Tran}, {Goodrich},  \& {Martel} 1999]{ocm+99}
{Ogle}, P.~M., {Cohen}, M.~H., {Miller}, J.~S., {Tran}, H.~D., {Goodrich},  R.~W., \& {Martel}, A.~R. 1999, \apjs, 125, 1

\bibitem[{Osmer}, {Porter}, \& {Green} 1994]{opg94}
{Osmer}, P.~S., {Porter}, A.~C., \& {Green}, R.~F. 1994, \apj, 436, 678

\bibitem[{Pei} 1992]{pei92}
{Pei}, Y.~C. 1992, \apj, 395, 130

\bibitem[{Pier et al.} 2002]{pmh+02}
{Pier et al.} 2002, \aj, submitted

\bibitem[{Press}, {Teukolsky}, {Vetterling}, \&  {Flannery} 1992]{ptv+92}
{Press}, W.~H., {Teukolsky}, S.~A., {Vetterling}, W.~T., \& {Flannery}, B.~P.  1992, {Numerical recipes in C. The art of scientific computing} (Cambridge:  University Press, |c1992, 2nd ed.)

\bibitem[{Prevot}, {Lequeux}, {Prevot}, {Maurice}, \&  {Rocca-Volmerange} 1984]{plp+84}
{Prevot}, M.~L., {Lequeux}, J., {Prevot}, L., {Maurice}, E., \&  {Rocca-Volmerange}, B. 1984, \aap, 132, 389

\bibitem[{Reichard et al.} 2003]{rrh+02}
{Reichard et al.} 2003, \aj, in preparation

\bibitem[{Richards}, {Fan}, {Newberg},  {Strauss}, {Vanden Berk}, {Schneider}, {Yanny}, {Boucher}, {Burles},  {Frieman}, {Gunn}, {Hall}, {Ivezi{\' c}}, {Kent}, {Loveday}, {Lupton},  {Rockosi}, {Schlegel}, {Stoughton}, {SubbaRao}, \& {York} 2002a]{rfn+02}
{Richards}, G.~T., {Fan}, X., {Newberg}, H.~J., {Strauss}, M.~A., {Vanden  Berk}, D.~E., {Schneider}, D.~P., {Yanny}, B., {Boucher}, A., {et al.} 2002a,  \aj, 123, 2945

\bibitem[{Richards}, {Fan}, {Schneider}, {Vanden  Berk}, {Strauss}, {York}, {Anderson}, {Tremonti}, {Uomoto}, {Waddell},  {Yanny}, \& {Zheng} 2001]{rfs+01}
{Richards}, G.~T., {Fan}, X., {Schneider}, D.~P., {Vanden Berk}, D.~E.,  {Strauss}, M.~A., {York}, D.~G., {Anderson}, J.~E., {Tremonti}, C., {et al.} 2001, \aj, 121, 2308

\bibitem[{Richards}, {Vanden Berk},  {Reichard}, {Hall}, {Schneider}, {SubbaRao}, {Thakar}, \& {York} 2002b]{ric+02b}
{Richards}, G.~T., {Vanden Berk}, D.~E., {Reichard}, T.~A., {Hall}, P.~B.,  {Schneider}, D.~P., {SubbaRao}, M., {Thakar}, A.~R., \& {York}, D.~G.  2002b, \aj, 124, 1

\bibitem[{Richards et al.} 2003]{rhv+02}
{Richards et al.} 2003, \aj, submitted

\bibitem[{Savage} \& {Mathis} 1979]{sm79}
{Savage}, B.~D. \& {Mathis}, J.~S. 1979, \araa, 17, 73

\bibitem[{Schlegel}, {Finkbeiner}, \&  {Davis} 1998]{sfd98}
{Schlegel}, D.~J., {Finkbeiner}, D.~P., \& {Davis}, M. 1998, \apj, 500, 525

\bibitem[{Schmidt} \& {Hines} 1999]{sh99}
{Schmidt}, G.~D. \& {Hines}, D.~C. 1999, \apj, 512, 125

\bibitem[{Schneider}, {Richards}, {Fan}, {Hall},  {Strauss}, {Vanden Berk}, {Gunn}, {Newberg}, {Reichard}, {Stoughton},  {Voges}, {Yanny}, {Anderson}, {Annis}, {Bahcall}, {Bauer}, {Bernardi},  {Blanton}, {Boroski}, {Brinkmann}, {Briggs}, {Brunner}, {Burles}, {Carey},  {Castander}, {Connolly}, {Csabai}, {Doi}, {Friedman}, {Saxe}, {Schlegel},  {Siegmund}, {Smee}, {Snir}, {SubbaRao}, {Szalay}, {Thakar}, {Uomoto},  {Waddell}, \& {York} 2002]{sch+02}
{Schneider}, D.~P., {Richards}, G.~T., {Fan}, X., {Hall}, P.~B., {Strauss},  M.~A., {Vanden Berk}, D.~E., {Gunn}, J.~E., {Newberg}, H.~J., {et al.} 2002, \aj, 123, 567

\bibitem[{Smith}, {Tucker}, {Kent}, {Richmond},  {Fukugita}, {Ichikawa}, {Ichikawa}, {Jorgensen}, {Uomoto}, {Gunn}, {Hamabe},  {Watanabe}, {Tolea}, {Henden}, {Annis}, {Pier}, {McKay}, {Brinkmann}, {Chen},  {Holtzman}, {Shimasaku}, \& {York} 2002]{stk+02}
{Smith}, J.~A., {Tucker}, D.~L., {Kent}, S., {Richmond}, M.~W., {Fukugita}, M.,  {Ichikawa}, T., {Ichikawa}, S., {Jorgensen}, A.~M., {et al.} 2002, \aj, 123, 2121

\bibitem[{Sprayberry} \& {Foltz} 1992]{sf92}
{Sprayberry}, D. \& {Foltz}, C.~B. 1992, \apj, 390, 39

\bibitem[{Stoughton}, {Lupton}, {Bernardi},  {Blanton}, {Burles}, {Castander}, {Connolly}, {Eisenstein}, {Frieman},  {Hennessy}, {Hindsley}, {Ivezi{\' c}}, {Kent}, {Kunszt}, {Lee}, {Meiksin},  {Munn}, {Newberg}, {Nichol}, {Nicinski}, {Pier}, {Richards}, {Richmond},  {Schlegel}, {Smith}, {Strauss}, {SubbaRao}, {Szalay}, {Thakar}, {Tucker},  {Vanden Berk}, {Yanny}, {Adelman}, {York}, {Zehavi}, \& {Zheng} 2002]{sto+02}
{Stoughton}, C., {Lupton}, R.~H., {Bernardi}, M., {Blanton}, M.~R., {Burles},  S., {Castander}, F.~J., {Connolly}, A.~J., {Eisenstein}, D.~J., {et al.} 2002, \aj, 123, 485

\bibitem[{Surdej} \& {Hutsemekers} 1987]{sh87}
{Surdej}, J. \& {Hutsemekers}, D. 1987, \aap, 177, 42

\bibitem[{Tolea}, {Krolik}, \& {Tsvetanov} 2002]{kro+02}
{Tolea}, A., {Krolik}, J.~H., \& {Tsvetanov}, Z. 2002, \apj, 578, L31

\bibitem[{Turnshek} 1984]{tur84}
{Turnshek}, D.~A. 1984, \apj, 280, 51

\bibitem[{Vanden Berk}, {Richards}, {Bauer},  {Strauss}, {Schneider}, {Heckman}, {York}, {Hall}, {Fan}, {Knapp}, {Yanny},  \& {Zheng} 2001]{vrb+01}
{Vanden Berk}, D.~E., {Richards}, G.~T., {Bauer}, A., {Strauss}, M.~A.,  {Schneider}, D.~P., {Heckman}, T.~M., {York}, D.~G., {Hall}, P.~B., {et al.} 2001, \aj, 122, 549

\bibitem[{Voges}, {Aschenbach}, {Boller}, {Br{\"  a}uninger}, {Briel}, {Burkert}, {Dennerl}, {Englhauser}, {Gruber}, {Haberl},  {Hartner}, {Hasinger}, {K{\" u}rster}, {Pfeffermann}, {Pietsch}, {Predehl},  {Rosso}, {Schmitt}, {Tr{\" u}mper}, \& {Zimmermann} 1999]{vab+99}
{Voges}, W., {Aschenbach}, B., {Boller}, T., {Br{\" a}uninger}, H., {Briel},  U., {Burkert}, W., {Dennerl}, K., {Englhauser}, J., {et al.} 1999, \aap, 349, 389

\bibitem[{Voit}, {Weymann}, \& {Korista} 1993]{vwk93}
{Voit}, G.~M., {Weymann}, R.~J., \& {Korista}, K.~T. 1993, \apj, 413, 95

\bibitem[{Weymann} 1995]{wey95}
{Weymann}, R.~J. 1995, in QSO Absorption Lines, Proceedings of the ESO Workshop  Held at Garching, Germany, 21 - 24 November 1994, edited by Georges Meylan.  Springer-Verlag Berlin Heidelberg New York. Also ESO Astrophysics Symposia,  1995., p.213, 213

\bibitem[{Weymann}, {Morris}, {Foltz}, \&  {Hewett} 1991]{wmf+91}
{Weymann}, R.~J., {Morris}, S.~L., {Foltz}, C.~B., \& {Hewett}, P.~C. 1991,  \apj, 373, 23

\bibitem[{White}, {Becker}, {Gregg},  {Laurent-Muehleisen}, {Brotherton}, {Impey}, {Petry}, {Foltz}, {Chaffee},  {Richards}, {Oegerle}, {Helfand}, {McMahon}, \& {Cabanela} 2000]{wbg+00}
{White}, R.~L., {Becker}, R.~H., {Gregg}, M.~D., {Laurent-Muehleisen}, S.~A.,  {Brotherton}, M.~S., {Impey}, C.~D., {Petry}, C.~E., {Foltz}, C.~B., {et al.} 2000, \apjs, 126, 133

\bibitem[{Wilkes} 1986]{wil86}
{Wilkes}, B.~J. 1986, \mnras, 218, 331

\bibitem[{Yamamoto} \& {Vansevi{\v c}ius} 1999]{yv99}
{Yamamoto}, T.~M. \& {Vansevi{\v c}ius}, V. 1999, \pasj, 51, 405

\bibitem[{York}, {Adelman}, {Anderson}, {Anderson},  {Annis}, {Bahcall}, {Bakken}, {Barkhouser}, {Bastian}, {Berman}, {Boroski},  {Bracker}, {Briegel}, {Briggs}, {Brinkmann}, {Brunner}, {Burles}, {Carey},  {Carr}, {Castander}, {Tucker}, {Uomoto}, {Vanden Berk}, {Vogeley}, {Waddell},  {Wang}, {Watanabe}, {Weinberg}, {Yanny}, \& {Yasuda} 2000]{yor+00}
{York}, D.~G., {Adelman}, J., {Anderson}, J.~E., {Anderson}, S.~F., {Annis},  J., {Bahcall}, N.~A., {Bakken}, J.~A., {Barkhouser}, R., {et al.} 2000, \aj, 120, 1579

\end{thebibliography}
\end{document}